\documentclass{aastex}
\usepackage{spr-astr-addons}
\usepackage{url}

\def\la{\mathrel{\hbox{\rlap{\hbox{\lower4pt\hbox{$\sim$}}}\hbox{$<$}}}}
\def\ga{\mathrel{\hbox{\rlap{\hbox{\lower4pt\hbox{$\sim$}}}\hbox{$>$}}}}
\def\lg{\mathrel{\hbox{\rlap{\hbox{\lower2.7pt\hbox{$>$}}}{\raise2.7pt\hbox{$<$}}}}}
\def\gl{\mathrel{\hbox{\rlap{\hbox{\lower2.7pt\hbox{$<$}}}{\raise2.7pt\hbox{$>$}}}}}
\def\half{{\textstyle{1\over2}}}
\def\Az{A \!\!\! | \,\,}
\def\Azp{A' \!\!\!\! | \,\,}
\def\ajp{Aust J.\ Phys.\ }
\def\ap{Ann.\ Phys.\ }
\def\aat{Astron.\ Astrophys.\ Trans.\ }
\def\ijmpa{Int.\ J.\ Mod.\ Phys.\ A\ }
\def\Pergamon{{Pergamon Press, Oxford}\ }
\def\Springer{{Springer, New York}\ }
\allowdisplaybreaks

\begin{document}
\title{Resonant Compton scattering associated with pair creation.}
\shorttitle{Resonant Compton scattering}
\shortauthors{J.I.Weise}
\author{Jeanette I Weise\altaffilmark{}}
\affil{School of Physics, University of Sydney, NSW 2006, Australia
                     email: \url{weise@physics.usyd.edu.au}}
%
\begin{abstract}
Studies of Compton scattering by relativistic electrons in a strong magnetic field have been restricted to either incident photon angles $\theta'$ aligned along the magnetic field $B$ or incident photon energies $\omega'$ below the first pair creation threshold $\omega'_{PC}$.  When these restrictions are relaxed there is a resonance in Compton scattering associated with pair creation (PC), that is analogous to but independent of known resonances associated with gyromagnetic absorption (GA).  As with the GA resonances, that may be labeled by the Landau quantum numbers of the relevant states, there is a sequence of PC resonances where the scattering cross section diverges.  In this paper, the lowest divergence is studied for incident photon energies satisfying ${\omega'}^2\sin^2\theta'/(2eB)\ll1$, assuming that the scattering electron is in its ground (Landau) state.  This lowest resonance affects only parallel-polarized photons.  
\end{abstract}
%
\section{Introduction}
Compton scattering of relativistic electrons in a strong magnetic field has been investigated for over four decades (eg:\cite{Ca71,H79,DH86,BAM86}).  Relatively recently \citet{G00} extended the theory to magnetic fields in excess of the critical magnetic field, $B_{\rm cr}=4.4\times10^{13}$G.  Such ultrastrong magnetic fields are relevant to magnetars which include soft gamma repeaters and anomalous X-ray pulsars (\cite{WT06,M08}). Gonthier et al. made several restrictive assumptions, including that the initial electron is in its ground (Landau) state, and that the initial photon is propagating parallel to the magnetic field. The first of these assumptions is well justified. An electron in a Landau state, $n>0$, relaxes in a very short time to its ground state, $n=0$, through gyromagnetic emission. The justification given for the second assumption can be summarized as follows: if photons are distributed over a wide range of angles in the laboratory frame, in which the electron is highly relativistic, then these photons are nearly all propagating at a small angle to the magnetic field in the electron's rest frame, by which is meant the frame in which its parallel momentum is zero, $p_z=0$. This small angle is assumed to be zero. However, the perpendicular momentum of the initial photon, which is denoted by $k'_\perp=\omega'\sin\theta'$ in natural units ($\hbar,\,c=1$), is an invariant, and so has the same value in both frames. Moreover, this invariant appears in the parameter $x'=k'^2_\perp/2eB$ in the theory, and setting $x'=0$ is overly restrictive, in that it excludes a resonance that is associated with pair creation (PC). This resonance is in addition to a resonance associated with gyromagnetic absorption (GA), which is well known in the pulsar context as resonant Compton scattering (\cite{DH86,D90,L96}). In this paper the assumption of parallel propagation for the initial photon is relaxed, and the scattering cross section including the PC resonance is considered.

\begin{figure}[t]
\begin{center}
\includegraphics[width=8cm]{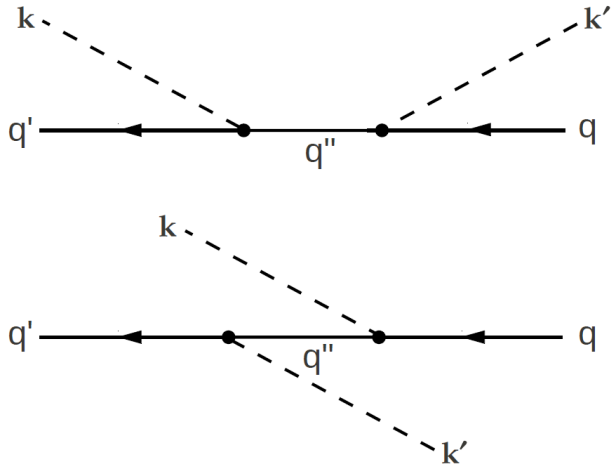}
\caption{The two Feynman diagrams for Compton scattering in a magnetic field which have different labeling of the particle states to those in the absence of the field.  For example, for the initial particle one has $q=n,p_z,\sigma$ where $n$ is the Landau quantum number, $p_z$ is the parallel momentum and $\sigma$ is the spin.  Both the energy and parallel momentum are conserved.}
\label{fig:Feyndiag}
\end{center}
\end{figure}

The two Feynman diagrams for Compton scattering are presented in Figure~\ref{fig:Feyndiag}, time increasing from right to left.  The solid lines represent particles (electrons or positrons) and the broken lines represent photons with wave 4-vectors ${k'}^\mu=(\omega',{\bf k}')$ and $k^\mu=(\omega,{\bf k})$.  Compton scattering is a second order process, depending on the fine structure constant squared, whereby an initial photon of energy $\omega'$ is absorbed by an initial particle with energy $\varepsilon$.  The particle jumps to an intermediate state of energy $\varepsilon''$ and subsequently deexcites to a final state of energy $\varepsilon'$ emitting a final photon of energy $\omega$.  The energy of the particle is made up of its rest energy, its parallel momentum (parallel to the magnetic field upon which the $z$-axis is aligned) and its perpendicular momentum which is a product of the magnetic field strength and the particle's Landau quantum number.  For the initial particle, this is of the form
\begin{equation}
\varepsilon=\sqrt{m^2\,+\,p_z^2\,+2neB}.
\end{equation}
The initial and final particles are taken to be electrons.  The intermediate particle can be either an electron or a positron and both these states are summed over as is their Landau quantum number $n''$.  The two types of resonance associated with Compton scattering appear as resonances in the first and the second Feynman amplitudes.  In the first Feynman diagram when the intermediate particle is an electron, the first type of resonance that occurs is the well known GA resonance.  Its divergence is usually handled by introducing a finite width, equal to the gyromagnetic decay width of the intermediate state, and adding it to the resonant denominator.  The PC resonance is associated with the second Feynman diagram, in which the intermediate particle is a positron.  Both of these resonances were identified as far back as 1979 by \citet{H79}, who restricted his calculations to photon energies below the first PC threshold.

At either a GA or a PC resonance the question arises as to whether the process should be regarded as a second-order process, corresponding to Compton scattering with a virtual intermediate state, or as a 2-step process, corresponding to two first-order processes with a real intermediate state. For the GA resonance, the first step is gyromagnetic absorption of the initial photon by the initial electron, with $n=0$, leaving an electron in an excited Landau level, $n''$. The second step is gyromagnetic emission of the final photon by this electron, with a transition back to the ground state $n'=0$. The lifetime of the intermediate state is determined by the (inverse of the) gyromagnetic decay probability. A conventional way of including this effect is through the natural line width, $\Gamma$, of the intermediate state, such that the resonant denominator is replaced according to $\omega+\varepsilon'-\varepsilon''\to\omega+\varepsilon'-\varepsilon''+i\Gamma/2$. For the PC resonance, the two steps are pair creation of the final electron and the intermediate positron by the initial photon, and annihilation of the initial electron and the intermediate positron to produce the final photon. In this case, the lifetime of the intermediate state is determined by the (inverse of the) rate of creation of an electron-positron pair by the initial photon.

At magnetic field strengths far in excess of $B_{cr}$, the effects of photon dispersion and wave function renormalisation become important for parallel-polarized photons (eg:\cite{HI13,SU10} and references therein).  This is clearly evident in the work by \citet{CR09} who investigated the Compton effect in a strongly magnetized ($B=200B_{cr}$) medium consisting of both vacuum and plasma contributions.  The purpose of the present investigation is to compare Compton scattering for a range of incident photon angles under the same conditions as employed by \citet{G00}, namely assuming the vacuum dispersion laws $k'^2=0$ and $k^2=0$ for the incident and final photons; Gonthier et al. only considered an initial photon angle $\theta'$ of $180^\circ$, defined as $0^\circ$ in their paper.  Although only over a limited range of photon energies and for incident and final electrons in their ground state, the results of the current work cover the entire range of incident photon angles.  The magnetic field strengths considered are $0.1B_{cr},\,B_{cr},\,10B_{cr}$ and $100B_{cr}$, the same as those chosen in \citet{G00}.  The notation used in this paper differs from that used by Gonthier et al. in that the incident photon's parameters are primed and the final photon's parameters are unprimed whereas in Gonthier et al.'s paper, they are the opposite.

The results found in the present paper show that, over the entire range of magnetic fields, the Compton scattering cross sections are greater in magnitude once the incident photon has a nonzero component of perpendicular momentum.  Further, at the two highest magnetic field strengths for which the first PC threshold falls within the range of the incident photon energies considered, the cross sections increase as the PC threshold is approached in a similar manner as when the first GA resonance is approached.  These results are relevant for the superstrong magnetic fields featured in magnetars, as discussed in \citet{G00}, where the Compton scattering process remains an important energy loss process.  For example, \citet{BH07}, \citet{N08} and \citet{Ba11} investigated resonant Compton scattering at the cyclotron frequencies as a mechanism behind the hard X-ray tails persistent in AXPs and magnetars. A related application is to the effect of Compton scattering on the shape of the radiation spectra of strongly magnetized neutron stars; for example, \citet{Mu12} derived the kinetic equations for Compton scattering as a basis for radiation transport models.  As discussed in this paper, the resonance at the first PC threshold for oblique incident photon angles and ultrastrong magnetic fields can be at a lower energy than the GA resonance, and give a previously unrecognized form of resonant Compton scattering in these applications.

The paper is set out as follows.  The relevant equations and approximations used in the evaluation of the Compton scattering cross section are set out in Sect.~2.  A detailed derivation of the second order $S$-matrix for Compton scattering from first principles is presented in this section.  The criterion for the intermediate state to be part of a 2-step process, rather that a virtual state in a scattering event, is discussed in Sect.~3.  In Sect.~4, specific details are given of how the PC resonance is handled.  The results are presented in Sect.~5 over a restricted region of incident photon energies below the first GA resonance.  Some discussion of the work by Gonthier et al. is given in this section.  In Sect.~6, a general summary of the important features of the paper is presented.  Natural units are used throughout.

\section{Compton scattering cross section}
In this section, the theory and derivation leading to the expressions for the Compton scattering cross sections are given.  This begins with the choice of wave functions, then the form of the $S$-matrix from which one obtains the probabilities and differential cross sections for each of the modes of polarization.
\subsection{Choice of wave functions}
Much of the theory in this section can be found in \citet{MP83}.  The wave solution to Dirac's equation in the presence of a magnetic field $B$ is of the form
\begin{equation}
\psi({\bf x},t)=f(x){\rm exp}(-i\epsilon\varepsilon_qt+i\epsilon p_yy+i\epsilon p_zz),
\label{solDir}
\end{equation}
where $\epsilon=\pm$ denotes both positive (electron) and negative (positron) energy solutions and the function $f(x)$ is a column matrix (Dirac spinor).  (Note $\epsilon$ denotes a sign and $\varepsilon$ denotes an energy.)  Introducing the Landau quantum number $n$ via
\begin{equation}
2neB\equiv\varepsilon^2_q-m^2-p^2_z,
\end{equation}
the solutions have
\begin{equation}
2n\mp1=2l+1,
\label{2nmp1}
\end{equation}
with the orbital quantum number $l=0,1,2,\dots$.  The ground state $(n=0)$ is nondegenerate with each excited state $(n=1,2,\dots)$ doubly degenerate.  The solutions are the normalised oscillator wave functions
\begin{equation}
v_l(\xi)=\frac{H_l(\xi){\rm exp}(-\half\xi^2)}{(\pi^{1/2}2^ll!)^{1/2}},
\end{equation}
where $H_l(\xi)$ is the Hermite polynomial and $\xi=\sqrt{eB}(x+\epsilon p_y/eB)$.  Hence one has
\begin{equation}
f(x)=\begin{bmatrix}C_1v_{n-1}(\xi) \\ C_2v_n(\xi) \\ C_3v_{n-1}(\xi) \\ C_4v_n(\xi)\end{bmatrix},
\end{equation}
where $C_1$ to $C_4$ are normalisation constants and $C_1=C_3=0$ when $n=0$.

From Eq.~(\ref{2nmp1}), one has
\begin{equation}
n=l+\half(\sigma+1)
\label{neql}
\end{equation}
where $\sigma=\pm1$ is interpreted as a spin eigenvalue.  The energy involves $n$ via
$$\varepsilon_q={\sqrt {m^2+p_z^2+2neB}},$$
such that Eq.~(\ref{neql}) implies a separation into an orbital part, described by $l$ and a spin part described by $\sigma$, respectively.  The energy states of the simple harmonic oscillator correspond to $(l+\half)\hbar\Omega_0$ with $\Omega_0=eB/m$ the frequency of the oscillator.  The remaining energy $\half\sigma\hbar\Omega_0$ represents the contribution $\bf \mu.B$ where $\bf \mu$ is the magnetic moment.  Hence the spin operator is chosen via the eigenstates of the magnetic moment operator of \citet{ST68}, viz.
\begin{eqnarray}
&&\begin{bmatrix}C_1 \\ C_2 \\ C_3 \\ C_4\end{bmatrix}={{\rm exp}\{i\Phi(\epsilon,\sigma)\}\over\sqrt{4\epsilon\sigma\varepsilon_q\varepsilon^0_q(\epsilon\varepsilon_q+\sigma\varepsilon^0_q)
(\sigma\varepsilon^0_q+m)}}\nonumber\\
&&\qquad\qquad\qquad\qquad\times\begin{bmatrix}(\epsilon\varepsilon_q+\sigma\varepsilon^0_q)(\sigma\varepsilon^0_q+m) \\ -i\epsilon p_zp_n \\ (\sigma\varepsilon^0_q+m)\epsilon p_z \\ i(\epsilon\varepsilon_q+\sigma\varepsilon^0_q)p_n\end{bmatrix},
\end{eqnarray}
where $\varepsilon^0_q=\sqrt{m^2+2neB}$, $p_n=\sqrt{2neB}$ and $\Phi(\epsilon,\sigma)$ is an arbitrary phase factor.  One choice of phase gives
\begin{eqnarray}
&&\psi^\epsilon_q({\bf x},t)={{\rm exp}(-i\epsilon\varepsilon_qt+i\epsilon p_yy+i\epsilon p_zz)\over\sqrt{4V\varepsilon_q\varepsilon^0_q(\varepsilon_q+\varepsilon^0_q)(\varepsilon^0_q+m)}}
\nonumber\\
&&\times\Bigg\{\delta_{\epsilon,+}\Bigg[\delta_{\sigma,+}\begin{bmatrix}(\varepsilon_q+\varepsilon^0_q)(\varepsilon^0_q+m)v_{n-1}(\xi) \\ -ip_zp_nv_n(\xi) \\ p_z(\varepsilon^0_q+m)v_{n-1}(\xi)\cr ip_n(\varepsilon_q+\varepsilon^0_q)v_n(\xi)\end{bmatrix}\nonumber\\
&&+\delta_{\sigma,-}\begin{bmatrix}-ip_zp_nv_{n-1}(\xi) \\ (\varepsilon_q+\varepsilon^0_q)(\varepsilon^0_q+m)v_n(\xi) \\ -ip_n(\varepsilon_q+\varepsilon^0_q)v_{n-1}(\xi) \\ -p_z(\varepsilon^0_q+m)v_n(\xi)\end{bmatrix}\Bigg]\nonumber\\
&&+\delta_{\epsilon,-}\Bigg[\delta_{\sigma,+}\begin{bmatrix}p_z(\varepsilon^0_q+m)v_{n-1}(\xi) \\ -ip_n(\varepsilon_q+\varepsilon^0_q)v_n(\xi) \\ (\varepsilon_q+\varepsilon^0_q)(\varepsilon^0_q+m)v_{n-1}(\xi) \\  ip_zp_nv_n(\xi)\end{bmatrix}\nonumber\\
&&+\delta_{\sigma,-}\begin{bmatrix}ip_n(\varepsilon_q+\varepsilon^0_q)v_{n-1}(\xi) \\ -p_z(\varepsilon^0_q+m)v_n(\xi) \\  ip_zp_nv_{n-1}(\xi) \\ (\varepsilon_q+\varepsilon^0_q)(\varepsilon^0_q+m)v_n(\xi)\end{bmatrix}\Bigg]\Bigg\},
\end{eqnarray}
where $V$ is the volume of the system.  

The choice of these wave functions, specifically the  the magnetic moment operator as the spin operator, is strongly preferred over the Johnson and Lippmann (JL) wave functions, used by \citet{G00}.  In particular, it is only for this choice of spin operator that the spin eigenvalue is a constant of the motion (\cite{ST68,H79,MP83}).   Further, the JL wave functions lack symmetry between the electron and positron states.

\subsection{Probability and modes of polarization}

Compton scattering is described by a probability (\cite{QPDII}), whose derivation using a $S$-matrix approach is summarized in Appendix~\ref{A-appendix}. The probability has the specific form
\begin{eqnarray}
&&w^{\epsilon'\epsilon}_{q'q}({\bf k},{\bf k}')={\mu^2_0e^4\over4\omega'\omega}\,2\pi\,\delta(\epsilon'\varepsilon'_{q'}-\epsilon\varepsilon_q-\omega'+\omega)
\nonumber\\
&&\qquad\qquad\times|e^*_{M\mu}({\bf k})e_{M'\nu}({\bf k}')[M^{\epsilon'\epsilon}_{q'q}({\bf k},{\bf k}')]^{\mu\nu}|^2,
\label{prob}
\end{eqnarray}
with $q''=n'',\,\epsilon''$ and
\begin{eqnarray}
&&[M^{\epsilon'\epsilon}_{q'q}({\bf k},{\bf k}')]^{\mu\nu}=[M^{\epsilon'\epsilon}_{q'q}({\bf k},{\bf k}')]_1^{\mu\nu}+[M^{\epsilon'\epsilon}_{q'q}({\bf k},{\bf k}')]_2^{\mu\nu},
\nonumber\\
&&[M^{\epsilon'\epsilon}_{q'q}({\bf k},{\bf k}')]_1^{\mu\nu}=\sum_{\epsilon'',q''}\,{[\Gamma^{\epsilon'\epsilon''}_{q'q''}({\bf k})]^\mu[\Gamma^{\epsilon\epsilon''}_{qq''}({\bf k}')]^{*\nu}\over 
\epsilon\varepsilon_q-\epsilon''\varepsilon''_{q''}+\omega'}
\nonumber\\
&&\qquad\qquad\times{\rm exp}\left(i{({\bf k\,\times\,k'})_z\over2eB}\right),
\nonumber\\
&&[M^{\epsilon'\epsilon}_{q'q}({\bf k},{\bf k}')]_2^{\mu\nu}=\sum_{\epsilon'',q''}\,{[\Gamma^{\epsilon''\epsilon}_{q''q}({\bf k})]^\mu[\Gamma^{\epsilon''\epsilon'}_{q''q'}({\bf k}')]^{*\nu}\over 
\epsilon\varepsilon_q-\epsilon''\varepsilon''_{q''}-\omega}
\nonumber\\
&&\qquad\qquad\times{\rm exp}\left(-i{({\bf k\,\times\,k'})_z\over2eB}\right).
\end{eqnarray}

\begin{figure}[t]
\begin{center}
\includegraphics[width=8cm]{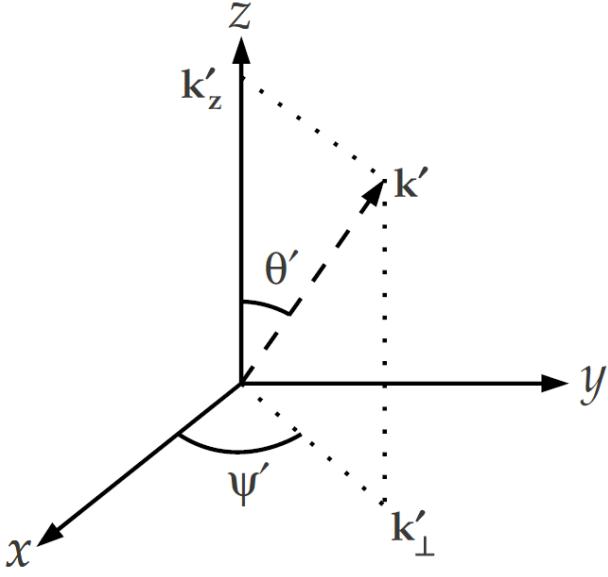}
\caption{The coordinate system adopted in this study for the incident photon in the rest frame of the initial electron.  The $z$-axis is chosen to be aligned along the magnetic field $\bf B$.  By replacing the primed parameters by unprimed parameters, the final photon is similarly represented.}
\label{fig:coordsys}
\end{center}
\end{figure}

The possible polarization modes of the initial and final photons are the two wave modes of the vacuum, denoted here as the perpendicular ($\perp$) and parallel ($\parallel$) polarized modes.  Consider the initial photon's wave vector ${\bf k}'$ in the coordinate system as shown in Figure~\ref{fig:coordsys}, with the $z$-axis aligned along the magnetic field.  The wave vector, making the angle $\theta'$ with the $z$-axis, has a component $k'_z(=\omega'\cos\theta')$ along the $z$-axis and a component in the $xy$-plane of $k'_\perp(=\omega'\sin\theta')$ at an angle $\psi'$ to the $x$-axis.  The perpendicular and parallel components of the polarization vector are defined as follows
\begin{eqnarray}
\perp':\ \ &&{\bf e'}\ \ {\rm along}\ \ -{\bf k'\times B},\nonumber\\
\parallel':\ \ &&{\bf e'}\ \ {\rm along}\ \ {\bf k'\times(k'\times B)}.
\label{primepol}
\end{eqnarray}
One is free to align the $x$-axis along the perpendicular component of the initial photon's wave vector ($\psi'=0$), so that Eq.~(\ref{primepol}) becomes
\begin{equation}
{\bf e'_\perp}=(0,1,0),\qquad\qquad{\bf e'_\parallel}=(\cos\theta',0,-\sin\theta').
\label{polprime}
\end{equation}
The final photon's wave vector, making an angle $\theta$ to the $z$-axis, has a component $k_z(=\omega\cos\theta)$ along the $z$-axis and a component in the $xy$-plane of $k_\perp(=\omega\sin\theta)$ at an angle $\psi$ to the $x$-axis, where $\psi$ in this case is arbitrary.  The polarization vectors for the final photon are thus
\begin{eqnarray}
&&{\bf e_\perp}=(-\sin\psi,\cos\psi,0),\nonumber\\
&&{\bf e_\parallel}=(\cos\theta\cos\psi,\cos\theta\sin\psi,-\sin\theta).
\label{pol}
\end{eqnarray}
The four possible polarization mode combinations for the initial and final photons for Compton scattering, hereafter referred to as the Compton scattering modes, are
$$\perp'\,\perp\,,\qquad\qquad\perp'\,\parallel\,,\qquad\qquad\parallel'\,\perp\,,\qquad\qquad\parallel'\,\parallel\,,$$
with the primed polarization referring to the initial photon and the unprimed the final photon.

The initial aim of the current work was to investigate the validity of setting $\theta'=180^\circ$, particularly as this choice of $\theta'$ effectively sets the Lorentz invariant associated with the perpendicular momentum,
\begin{equation}
p'_{\perp\gamma}\ =\ p'_{\perp\gamma{\rm lab}}\ \Longrightarrow\ \omega'\sin\theta'=\omega'_{\rm lab}\sin\theta'_{\rm lab},
\end{equation}
to zero, where the subscript ``lab'' identifies the laboratory frame.  Further the initial photons that take part in the scattering process may not be isotropic in the laboratory frame.  For example, if the photons are those associated with curvature radiation, then they will be initially at a $\theta'_{\rm lab}$ angle of approximately one over the Lorentz factor, $\gamma$.  A very small range of $\theta'_{\rm lab}$ between zero and $1/\gamma$ transforms to a large range of $\theta'$, namely between $180^\circ$ and $90^\circ$, in the rest frame of the initial electron.

As soon as $\theta'$ is no longer $180^\circ$, the evaluation becomes complicated with infinite sums over $n''$, the Landau quantum number of the intermediate particle state.   One means of truncating these infinite sums is to limit the arguments $x'$ and $x$ of the $J$-functions, that are inherent in the matrix elements, to small values, viz.
\begin{equation}
x'={{k'_\perp}^2\over2eB}\ll1,\qquad\qquad x={k_\perp^2\over2eB}\ll1.
\label{xxpineq}
\end{equation}
This effectively allows one to consider just the first few $n''$ values.  In addition, by including higher $n''$ values, one can include higher order terms in $x'$ and $x$.  In Gonthier et al.'s evaluation, $x'$ was zero.  To decide whether one has a true scattering event or a 2-step process, the decay widths for pair creation, $\Gamma_{\rm PC}$, are evaluated using Sokolov and Ternov wave functions and under the specific conditions that are current in this analysis.  These conditions and the resultant decay rates are summarized in Appendix~\ref{B-appendix}.   When the initial and final particles are electrons ($\epsilon,\,\epsilon'=+$) and the initial particle is in its ground state ($n=0$, $\sigma=-1$, with $p_z=0$), the expression for the probability of Compton scattering in Eq.~(\ref{prob}) simplifies considerably.  Further, the sum over $\sigma''$ is performed with $\epsilon''p''_z=\epsilon p_z+k'_z\equiv k'_z$ in the matrix element $[M^{\epsilon'\epsilon}_{q'q}({\bf k},{\bf k}')]_1^{\mu\nu}$ and $\epsilon''p''_z=\epsilon p_z-k_z\equiv -k_z$ in the matrix element $[M^{\epsilon'\epsilon}_{q'q}({\bf k},{\bf k}')]_2^{\mu\nu}$.  Choosing $\psi'=0$, the result after summing over $\sigma''$ is
\begin{eqnarray}
&&w^{+\,+}_{n'\,0}=\ \ {\mu_0^2e^4\over4\omega\omega'}\,{2\pi\,\delta(\varepsilon'-m-\omega'+\omega)\over 
16\varepsilon'{\varepsilon'}_0^0(\varepsilon'+{\varepsilon'}_0^0)({\varepsilon'}_0^0+m)}
\nonumber\\
&&\bigg|e_{M\,i}^*e'_{M'\,j}\,\bigg\{e^{i(xx')^{1/2}\sin\psi}\,\sum_{n''}\,
[c^{+\,+}_{n'\,0}]^{ij}_{n''}e^{-i\psi(n'-n'')}
\nonumber\\
&&+e^{-i(xx')^{1/2}\sin\psi}\,\sum_{n''}\,
[d^{+\,+}_{n'\,0}]^{ij}_{n''}e^{-in''\psi}\bigg\}\bigg|^2.
\label{probpp}
\end{eqnarray}

The matrix elements $[c^{+\,+}_{n'\,0}]^{ij}_{n''}$ and $[d^{+\,+}_{n'\,0}]^{ij}_{n''}$ involve sums over $\epsilon''=\pm$.  In $[c^{+\,+}_{n'\,0}]^{ij}_{n''}$, when the intermediate particle is an electron ($\epsilon''=+$), the energy denominator, $\varepsilon'-\varepsilon''+\omega\ (=m-\varepsilon''+\omega')$, includes the GA resonance.  When the intermediate particle is a positron ($\epsilon''=-$), this denominator becomes $\varepsilon'+\varepsilon''+\omega\ (=m+\varepsilon''+\omega')$, which does not have a zero (for $\omega'>0$).  In $[d^{+\,+}_{n'\,0}]^{ij}_{n''}$, when the intermediate particle is a positron ($\epsilon''=-$), the energy denominator, $\varepsilon'+\varepsilon''-\omega'\ (=m+\varepsilon''-\omega)$, does have a zero and this corresponds to the PC resonance.  When the intermediate particle is an electron ($\epsilon''=+$), this denominator becomes $m-\varepsilon''-\omega\ (=\varepsilon'-\varepsilon''-\omega')$, which does not have a zero.  It represents the non-resonant part.

The matrix element $[c^{+\,+}_{n'\,0}]^{ij}_{n''}$ is as follows:
\begin{eqnarray}
&&[c^{+\,+}_{n'\,0}]^{ij}_{n''}=(-)^{n'}{1\over4\varepsilon''\sqrt{\varepsilon'\varepsilon'_0(\varepsilon'+\varepsilon'_0)(\varepsilon'_0+m)}}\nonumber\\
&&\qquad\times\sum_{\epsilon''=\pm}\,{\epsilon''\over\varepsilon'-\epsilon''\varepsilon''+\omega}
\nonumber\\
&&\times\Big\{\delta_{\sigma',+}\big\{-[A_1(\epsilon'',+)p_{n'}\,J^{n'}_{n''-n'}(x)\,J^0_{n''}(x')\nonumber\\
&&\ +B_{1,2}(\epsilon'',+)p_{n''}\,J^{n'-1}_{n''-n'}(x)\,J^0_{n''}(x')]b^ib^j
\nonumber\\
&&+[A_2(\epsilon'',+)p_{n'}e^{-i\psi}\,J^{n'}_{n''-n'-1}(x)\,J^0_{n''-1}(x')e^i_+e^j_-\nonumber\\
&&\ +B_{1,2}(\epsilon'',+)p_{n''}e^{i\psi}\,J^{n'-1}_{n''-n'+1}(x)\,J^0_{n''-1}(x')e^i_-e^j_-]
\nonumber\\
&&+[A_3(\epsilon'',+)e^{i\psi}\,J^{n'-1}_{n''-n'+1}(x)\,J^0_{n''}(x')e^i_-b^j\nonumber\\
&&\ -B_{3,4}(\epsilon'',+)p_{n'}p_{n''}e^{-i\psi}\,J^{n'}_{n''-n'-1}(x)\,J^0_{n''}(x')e^i_+b^j]
\nonumber\\
&&+[A_4(\epsilon'',+)\,J^{n'-1}_{n''-n'}(x)\,J^0_{n''-1}(x')\nonumber\\
&&\ -B_{3,4}(\epsilon'',+)p_{n'}p_{n''}\,J^{n'}_{n''-n'}(x)\,J^0_{n''-1}(x')]b^ie^j_-\big\}
\nonumber\\
&&+\delta_{\sigma',-}\big\{[A_1(\epsilon'',-)\,J^{n'}_{n''-n'}(x)\,J^0_{n''}(x')\nonumber\\
&&\ -B_{1,2}(\epsilon'',-)p_{n'}p_{n''}\,J^{n'-1}_{n''-n'}(x)\,J^0_{n''}(x')]b^ib^j
\nonumber\\
&&-[A_2(\epsilon'',-)e^{-i\psi}\,J^{n'}_{n''-n'-1}(x)\,J^0_{n''-1}(x')e^i_+e^j_-\nonumber\\
&&\ -B_{1,2}(\epsilon'',-)p_{n'}p_{n''}e^{i\psi}\,J^{n'-1}_{n''-n'+1}(x)\,J^0_{n''-1}(x')e^i_-e^j_-]
\nonumber\\
&&+[A_3(\epsilon'',-)p_{n'}e^{i\psi}\,J^{n'-1}_{n''-n'+1}(x)\,J^0_{n''}(x')e^i_-b^j\nonumber\\
&&\ +B_{3,4}(\epsilon'',-)p_{n''}e^{-i\psi}\,J^{n'}_{n''-n'-1}(x)\,J^0_{n''}(x')e^i_+b^j]
\nonumber\\
&&+[A_4(\epsilon'',-)p_{n'}\,J^{n'-1}_{n''-n'}(x)\,J^0_{n''-1}(x')\nonumber\\
&&\ +B_{3,4}(\epsilon'',-)p_{n''}\,J^{n'}_{n''-n'}(x)\,J^0_{n''-1}(x')]b^ie^j_-\big\}\Big\},
\label{cnp0}
\end{eqnarray}
where ${\bf e}_\pm=(1,\pm i,0),\ {\bf b}=(0,0,1)$, and
\begin{eqnarray}
&&A_1(\epsilon'',+)=A_4(\epsilon'',-)\nonumber\\
&&\qquad=k'_z(\varepsilon'+\varepsilon'_0)+(k'_z-k_z)(\epsilon''\varepsilon''-m),
\nonumber\\
&&B_{1,2}(\epsilon'',+)=B_{3,4}(\epsilon'',-)=(k'_z-k_z)(\varepsilon'_0+m),
\nonumber\\    
&&A_2(\epsilon'',+)=A_3(\epsilon'',-)\nonumber\\
&&\qquad=k'_z(\varepsilon'+\varepsilon'_0)-(k'_z-k_z)(\epsilon''\varepsilon''-m),
\nonumber\\
&&A_3(\epsilon'',+)=A_2(\epsilon'',-)\nonumber\\
&&\qquad=(k'_z(k'_z-k_z)-(\varepsilon'+\varepsilon'_0)(\epsilon''\varepsilon''-m))(\varepsilon'_0+m),
\nonumber\\
&&B_{3,4}(\epsilon'',+)=B_{1,2}(\epsilon'',-)=(\varepsilon'+\varepsilon'_0),
\nonumber\\    
&&A_4(\epsilon'',+)=A_1(\epsilon'',-)\nonumber\\
&&\ =(k'_z(k'_z-k_z)+(\varepsilon'+\varepsilon'_0)(\epsilon''\varepsilon''-m))(\varepsilon'_0+m).
\end{eqnarray}
Since $p''_z$ only appears in $\varepsilon''$ and then as ${p_z''}^2$, it has the value $k'_z$ irrespective of the sign $\epsilon''$.

The matrix element $[d^{+\,+}_{n'\,0}]^{ij}_{n''}$ is as follows:
\begin{eqnarray}
&&[d^{+\,+}_{n'\,0}]^{ij}_{n''}={1\over4\varepsilon''\sqrt{\varepsilon'\varepsilon'_0(\varepsilon'+\varepsilon'_0)(\varepsilon'_0+m)}}\nonumber\\
&&\qquad\times\sum_{\epsilon''=\pm}\,{\epsilon''\over\varepsilon'-\epsilon''\varepsilon''-\omega'}
\nonumber\\
&&\times\Big\{\delta_{\sigma',+}\big\{[A'_1(\epsilon'',+)p_{n'}\,J^0_{n''}(x)\,J^{n'}_{n''-n'}(x')\nonumber\\
&&\ -B'_{1,2}(\epsilon'',+)p_{n''}\,J^0_{n''}(x)\,J^{n'-1}_{n''-n'}(x')]b^ib^j
\nonumber\\
&&-[A'_2(\epsilon'',+)p_{n'}e^{i\psi}\,J^0_{n''-1}(x)\,J^{n'}_{n''-n'-1}(x')e^i_-e^j_+\nonumber\\
&&\ -B'_{1,2}(\epsilon'',+)p_{n''}e^{i\psi}\,J^0_{n''-1}(x)\,J^{n'-1}_{n''-n'+1}(x')e^i_-e^j_-]
\nonumber\\
&&+[A'_3(\epsilon'',+)e^{i\psi}\,J^0_{n''-1}(x)\,J^{n'-1}_{n''-n'}(x')\nonumber\\
&&\ +B'_{3,4}(\epsilon'',+)p_{n'}p_{n''}e^{i\psi}\,J^0_{n''-1}(x)\,J^{n'}_{n''-n'}(x')]e^i_-b^j
\nonumber\\
&&+[A'_4(\epsilon'',+)\,J^0_{n''}(x)\,J^{n'-1}_{n''-n'+1}(x')b^ie^j_-\nonumber\\
&&\ +B'_{3,4}(\epsilon'',+)p_{n'}p_{n''}\,J^0_{n''}(x)\,J^{n'}_{n''-n'-1}(x')b^ie^j_+]\big\}
\nonumber\\
&&+\delta_{\sigma',-}\big\{-[A'_1(\epsilon'',-)\,J^0_{n''}(x)\,J^{n'}_{n''-n'}(x')\nonumber\\
&&\ +B'_{1,2}(\epsilon'',-)p_{n'}p_{n''}\,J^0_{n''}(x)\,J^{n'-1}_{n''-n'}(x')]b^ib^j
\nonumber\\
&&+[A'_2(\epsilon'',-)e^{i\psi}\,J^0_{n''-1}(x)\,J^{n'}_{n''-n'-1}(x')e^i_-e^j_+\nonumber\\
&&\ +B'_{1,2}(\epsilon'',-)p_{n'}p_{n''}e^{i\psi}\,J^0_{n''-1}(x)\,J^{n'-1}_{n''-n'+1}(x')e^i_-e^j_-]
\nonumber\\
&&+[A'_3(\epsilon'',-)p_{n'}e^{i\psi}\,J^0_{n''-1}(x)\,J^{n'-1}_{n''-n'}(x')\nonumber\\
&&\ -B'_{3,4}(\epsilon'',-)p_{n''}e^{i\psi}\,J^0_{n''-1}(x)\,J^{n'}_{n''-n'}(x')]e^i_-b^j
\nonumber\\
&&+[A'_4(\epsilon'',-)p_{n'}\,J^0_{n''}(x)\,J^{n'-1}_{n''-n'+1}(x')b^ie^j_-\nonumber\\
&&\ -B'_{3,4}(\epsilon'',-)p_{n''}\,J^0_{n''}(x)\,J^{n'}_{n''-n'-1}(x')b^ie^j_+]\big\}\Big\},
\label{dnp0}
\end{eqnarray}
where
\begin{eqnarray}
&&A'_1(\epsilon'',+)=A'_3(\epsilon'',-)\nonumber\\
&&\qquad=k_z(\varepsilon'+\varepsilon'_0)-(k'_z-k_z)(\epsilon''\varepsilon''-m),
\nonumber\\
&&B'_{1,2}(\epsilon'',+)=B'_{3,4}(\epsilon'',-)=(k'_z-k_z)(\varepsilon'_0+m),
\nonumber\\    
&&A'_2(\epsilon'',+)=A'_4(\epsilon'',-)\nonumber\\
&&\qquad=k_z(\varepsilon'+\varepsilon'_0)+(k'_z-k_z)(\epsilon''\varepsilon''-m),
\nonumber\\
&&A'_3(\epsilon'',+)=A'_1(\epsilon'',-)\nonumber\\
&&\qquad=(k_z(k'_z-k_z)-(\varepsilon'+\varepsilon'_0)(\epsilon''\varepsilon''-m))(\varepsilon'_0+m),
\nonumber\\
&&B'_{3,4}(\epsilon'',+)=B'_{1,2}(\epsilon'',-)=(\varepsilon'+\varepsilon'_0),
\nonumber\\    
&&A'_4(\epsilon'',+)=A'_2(\epsilon'',-)\nonumber\\
&&\ =(k_z(k'_z-k_z)+(\varepsilon'+\varepsilon'_0)(\epsilon''\varepsilon''-m))(\varepsilon'_0+m).
\end{eqnarray}
Since $p''_z$ only appears in $\varepsilon''$ and then as ${p_z''}^2$, it has the value $k_z$ irrespective of the value of $\epsilon''$.

\subsection{Differential cross section}  
For comparison purposes with previous work, it is convenient to express the Compton scattering probability as a differential cross section.  The relationship between the two is (\cite{MS72}):
\begin{eqnarray}
&&{d\sigma_{MM'}\over d\Omega}=\int_0^\infty d\omega{\omega^3n^2_M\over(2\pi)^3\omega'v'_{gM'}}{\partial(\omega n_M)\over\partial\omega}{\partial\cos\theta\over\partial\cos\theta_r}
\nonumber\\
&&\qquad\qquad\qquad\times {\partial\cos\theta'\over\partial\cos\theta'_r}\ w^{+\,+}_{q'\,q},
\end{eqnarray}
with $w^{+\,+}_{q'\,q}$ as given in Eq.~(\ref{probpp}) and where $n_M$ is the refractive index, $v'_{gM'}$ is a group velocity, and, $\theta_r$ and $\theta'_r$ are the ray angles of the scattered and incident photons respectively.  Taking $n_M$, $\partial(\omega n_M)/\partial\omega$, $\partial\cos\theta/\partial\cos\theta_r$, $\partial\cos\theta'/\partial\cos\theta'_r$ as unity and $v'_{gM'}$ as $c$, the integral over $\omega$ is performed over the $\delta$-function in the equation for $w^{+\,+}_{q'\,q}$ to give
\begin{eqnarray}
&&\int d\omega\ \delta(\omega-\omega'-m+\varepsilon')=\nonumber\\
&&\qquad\qquad{\omega'-\omega+m\over m-\omega\sin^2\theta+\omega'(1-\cos\theta'\cos\theta)},
\end{eqnarray}
the numerator of which is equal to $\varepsilon'$ by conservation of energy.  In terms of the Thomson scattering cross section,
$$\sigma_T={8\pi\over3}\left({\mu_0e^2\over4\pi m}\right)^2,$$
one obtains
\begin{eqnarray}
&&{d\sigma\over d\Omega}={3\sigma_T\over8\pi}\,{m^2\omega^2\over{\omega'}^2}{\omega'-\omega+m\over(m-\omega\sin^2\theta+\omega'(1-\cos\theta\cos\theta'))}
\nonumber\\
&&\times\bigg|e_{M\,i}^*e'_{M'\,j}\,\bigg\{e^{-i(xx')^{1/2}\sin\psi}\,\sum_{n''}\,
[c^{+\,+}_{n'\,0}]^{ij}_{n''}e^{-i\psi(n'-n'')}\nonumber\\
&&\ +e^{i(xx')^{1/2}\sin\psi}\,\sum_{n''}\,
[d^{+\,+}_{n'\,0}]^{ij}_{n''}e^{-in''\psi}\bigg\}\bigg|^2,
\label{dsdO}
\end{eqnarray}
where $d\Omega=d\cos\theta d\psi$.  The energy of the final photon must satisfy $\omega\leq\omega'$.  Specifically, for given $\omega',\,\theta',\,\theta,\,B,\,n'$ and using the conservation of energy and parallel momentum equations in Eq.~(\ref{conserve}), $\omega$ is given by
\begin{equation}
\omega={R\over T+\sqrt{T^2-R\sin^2\theta}},
\label{wfromwp}
\end{equation}
where $R={k'_\perp}^2+2m\omega'-2n'eB$ and $T=\omega'-k'_z\cos\theta+m$.

The simplest case and that studied in this work is for $n'=0$, whereby only the non spin-flip ($\sigma'=-$) part is allowed such that Eqs.~(\ref{cnp0}) and (\ref{dnp0}) simplify to 
\begin{eqnarray}
&&[c^{+\,+}_{0\,0}]^{ij}_{n''}={1\over4\sqrt{\varepsilon'm(\varepsilon'+m)2m}}
\nonumber\\
&&\times\Big\{C_1J^0_{n''}(x)J^0_{n''}(x')b^ib^j\nonumber\\
&&\qquad\qquad-C_2\,e^{-i\psi}J^0_{n''-1}(x)J^0_{n''-1}(x')e^i_+e^j_-
\nonumber\\
&&+C_3\,p_{n''}\Big[e^{-i\psi}J^0_{n''-1}(x)J^0_{n''}(x')e^i_+b^j\nonumber\\
&&\qquad\qquad+J^0_{n''}(x)J^0_{n''-1}(x')b^ie^j_-\Big]\Big\},
\label{c00}
\end{eqnarray}
and
\begin{eqnarray}
&&[d^{+\,+}_{0\,0}]^{ij}_{n''}={1\over4\sqrt{\varepsilon'm(\varepsilon'+m)2m}}
\nonumber\\
&&\times\Big\{-D_1J^0_{n''}(x)J^0_{n''}(x')b^ib^j\nonumber\\
&&\qquad\qquad+D_2\,e^{i\psi}J^0_{n''-1}(x)J^0_{n''-1}(x')e^i_-e^j_+
\nonumber\\
&&-D_3\,p_{n''}e^{i\psi}J^0_{n''-1}(x)J^0_{n''}(x')e^i_-b^j\nonumber\\
&&\qquad\qquad-D_3\,p_{n''}J^0_{n''}(x)J^0_{n''-1}(x')b^ie^j_+\Big\},
\label{d00}
\end{eqnarray}
where
\begin{eqnarray}
&&C_1\,=\,\sum_{\epsilon''=\pm}\,{\epsilon''A_1(\epsilon'',-)\over\varepsilon''(\varepsilon'-\epsilon''\varepsilon''+\omega)}\nonumber\\
&&\qquad={2(\varepsilon'_0+m)\{k'_z(k'_z-k_z)+\omega'(\varepsilon'+\varepsilon'_0)\}\over(\varepsilon'+\omega)^2-{\varepsilon''}^2},
\nonumber\\
&&C_2\,=\,\sum_{\epsilon''=\pm}\,{\epsilon''A_2(\epsilon'',-)\over\varepsilon''(\varepsilon'-\epsilon''\varepsilon''+\omega)}\nonumber\\
&&\qquad={2(\varepsilon'_0+m)\{k'_z(k'_z-k_z)-\omega'(\varepsilon'+\varepsilon'_0)\}\over(\varepsilon'+\omega)^2-{\varepsilon''}^2},
\nonumber\\
&&C_3\,=\,\sum_{\epsilon''=\pm}\,{\epsilon''B_{3,4}(\epsilon'',-)\over\varepsilon''(\varepsilon'-\epsilon''\varepsilon''+\omega)}\nonumber\\
&&\qquad={2(\varepsilon'_0+m)(k'_z-k_z)\over(\varepsilon'+\omega)^2-{\varepsilon''}^2},
\nonumber\\
&&D_1\,=\,\sum_{\epsilon''=\pm}\,{\epsilon''A'_1(\epsilon'',-)\over\varepsilon''(\varepsilon'-\epsilon''\varepsilon''-\omega')}\nonumber\\
&&\qquad={2(\varepsilon'_0+m)\{k_z(k'_z-k_z)+\omega(\varepsilon'+\varepsilon'_0)\}\over(\varepsilon'-\omega)^2-{\varepsilon''}^2},
\nonumber\\
&&D_2\,=\,\sum_{\epsilon''=\pm}\,{\epsilon''A'_2(\epsilon'',-)\over\varepsilon''(\varepsilon'-\epsilon''\varepsilon''-\omega')}\nonumber\\
&&\qquad={2(\varepsilon'_0+m)\{k_z(k'_z-k_z)-\omega(\varepsilon'+\varepsilon'_0)\}\over(\varepsilon'-\omega)^2-{\varepsilon''}^2},
\nonumber\\
&&D_3\,=\,\sum_{\epsilon''=\pm}\,{\epsilon''B'_{3,4}(\epsilon'',-)\over\varepsilon''(\varepsilon'-\epsilon''\varepsilon''-\omega')}\nonumber\\
&&\qquad={2(\varepsilon'_0+m)(k'_z-k_z)\over(\varepsilon'-\omega)^2-{\varepsilon''}^2}.
\label{CD}
\end{eqnarray}
Away from the resonances, the sums over $\epsilon''$ can be performed to give the second expressions in Eq.~(\ref{CD}), where the denominator in $C_1,\,C_2$ and $C_3$ can be simplified to ${k'_\perp}^2+2m\omega'-2n''eB$, and the denominator in $D_1,\,D_2$ and $D_3$ can be simplified to $k_\perp^2-2m\omega-2n''eB$ using the conservation of energy equation for Compton scattering, viz. $m+\omega'=\varepsilon'+\omega$, and $\varepsilon''=\sqrt{m^2+k_z^2+2n''eB}$.  Close to the resonances, where decay widths may need to be incorporated in the resonant parts $\epsilon''=+$ in $C_i,\ i=1,2,3$ and $\epsilon''=-$ in $D_1$, the resonant and non-resonant parts are treated separately and the first expressions are used.  The first terms in Eqs.~(\ref{c00}) and (\ref{d00}), namely those terms with $C_1$ and $D_1$, contribute to the $\parallel'\parallel$ mode only, whereby $n''\geq0$.  Of these two terms, only $D_1$ has the resonant denominator at $\epsilon''=-$ and $\omega'=\omega'_{PC}$.  The second terms contribute to all the modes and have $n''\geq1$; it is the only contribution to the $\perp'\perp$ mode.  The third terms contribute to the $\parallel'\perp$ and $\parallel'\parallel$ modes only and have $n''\geq1$, and the fourth terms contribute to the $\perp'\parallel$ and $\parallel'\parallel$ modes only and also have $n''\geq1$.

The resulting differential cross sections are as follows.  When $\theta'=180^\circ$ or $0^\circ$, one has $x'=0$ and $n''=1$ and
\begin{eqnarray}
&&{d\sigma^{\perp'\perp}\over d\cos\theta}={d\sigma^{\parallel'\perp}\over d\cos\theta}={3\sigma_T\over4}\,{\omega^2\over{\omega'}^2}\nonumber\\
&&\times{e^{-x}\over(m-\omega\sin^2\theta+\omega'+\omega'\cos\theta)}\,{\left\{(C_2)^2+(D_2)^2\right\}\over32(\varepsilon'+m)},
\nonumber\\
&&{d\sigma^{\perp'\parallel}\over d\cos\theta}={d\sigma^{\parallel'\parallel}\over d\cos\theta}={3\sigma_T\over4}\,{\omega^2\over{\omega'}^2}\nonumber\\
&&\times{e^{-x}\omega^2\over(m-\omega\sin^2\theta+\omega'+\omega'\cos\theta)}{1\over32(\varepsilon'+m)}
\nonumber\\
&&\times\Big\{\cos^2\theta\left[(C_2)^2+(D_2)^2\right]\nonumber\\
&&\qquad+\omega^2\sin^4\theta\left[(C_3)^2+(D_3)^2\right]\nonumber\\
&&\qquad+2\omega\sin^2\theta\cos\theta\left[(C_2)(C_3)+(D_2)(D_3)\right]\Big\}.
\end{eqnarray}
As $n''\neq0$, there is no PC divergence and hence no PC correction is necessary.  The results above should be the same as those obtained by \citet{G00}.  However in deriving the expression for the differential cross section from the probability (Eq.~(\ref{prob})), an error was discovered in the multiplicative factor in the cross section as derived by Gonthier et al.  They have a frequency factor that varies linearly as the final photon's energy over the initial photon's energy, whereas it varies quadratically with this ratio (see Eq.~(\ref{dsdO})).  This leads to an overestimation of their cross section particularly above the cyclotron resonance (see Sect.~5).  In the current evaluation, the eigenfunctions of the magnetic moment operator introduced by Sokolov and Ternov are used instead of the Johnson-Lippmann wave functions that Gonthier et al. used.  As the spin states are summed over, the two results should coincide, which they do for $\theta'=180^\circ$ and the correct frequency factor.

When $\theta'\neq180^\circ,\,0^\circ$, one has $0\leq n''\leq n_{max}+1$, where $n_{max}$ indicates the order of the expansion in $x'$.  The three modes $\perp'\perp,\,\perp'\parallel$ and $\parallel'\perp$, with $n''\geq1$, do not exhibit the PC divergence which is associated with $n''=0$.  The differential cross sections for these three modes are
\begin{eqnarray}
&&{d\sigma^{\perp'\perp}\over d\cos\theta}={3\sigma_T\over4}\,{e^{-(x+x')}\omega^2\over32\,{\omega'}^2(\varepsilon'+m)}
\nonumber\\
&&\times{1\over(m-\omega\sin^2\theta+\omega'-\omega'\cos\theta'\cos\theta)}\nonumber\\
&&\times\Big\{\sum^{n_{max}+1}_{n''=1}J_0\left[(C_2)^2+(D_2)^2\right]\nonumber\\
&&-2\sum^{n_{max}}_{n''=1}\sum^{n_{max}-n''+1}_{n=1}J_1{\cal B}\,(C_2)_{n''}(D_2)_n\Big\},
\nonumber\\
&&{d\sigma^{\perp'\parallel}\over d\cos\theta}={3\sigma_T\over4}{e^{-(x+x')}\omega^2\over32{\omega'}^2(\varepsilon'+m)}
\nonumber\\
&&\times{1\over(m-\omega\sin^2\theta+\omega'-\omega'\cos\theta'\cos\theta)}\nonumber\\
&&\times\Bigg\{\cos^2\theta\Big[\sum_{n''=1}J_0\left[(C_2)^2+(D_2)^2\right]\,\nonumber\\
&&+2\sum_{n''=1}\sum_{n=1}J_1{\cal B}\,(C_2)_{n''}(D_2)_n\Big]\nonumber\\
&&+\omega^2\sin^4\theta\Big[\sum_{n''=1}\,J_0\,\left[(C_3)^2+(D_3)^2\right]\nonumber\\
&&+2\sum_{n''=1}\sum_{n=1}J_1{\cal B}\,(C_3)_{n''}(D_3)_n\Big]\nonumber\\
&&+2\omega\cos\theta\sin^2\theta\Big[\sum_{n''=1}\,J_0\,\left[(C_2)(C_3)+(D_2)(D_3)\right]\nonumber\\
&&+\sum_{n''=1}\sum_{n=1}J_1{\cal B}\left[(C_2)_{n''}(D_3)_n+(C_3)_{n''}(D_2)_n\right]\Big]\Bigg\},
\nonumber\\
&&{d\sigma^{\parallel'\perp}\over d\cos\theta}={3\sigma_T\over4}{e^{-(x+x')}\omega^2\over32\,{\omega'}^2(\varepsilon'+m)}
\nonumber\\
&&\times{1\over(m-\omega\sin^2\theta+\omega'-\omega'\cos\theta'\cos\theta)}\nonumber\\
&&\times\Bigg\{\cos^2\theta'\Big[\sum_{n''=1}J_0\left[(C_2)^2+(D_2)^2\right]\nonumber\\
&&+2\sum_{n''=1}\sum_{n=1}J_1{\cal B}(C_2)_{n''}(D_2)_n\Big]\nonumber\\
&&+{\omega'}^2\sin^4\theta'\Big[\sum_{n''=1}J_0\left[(C_3)^2+(D_3)^2\right]\nonumber\\
&&+2\sum_{n''=1}\sum_{n=1}J_1{\cal B}\,(C_3)_{n''}(D_3)_n\Big]\nonumber\\
&&+2\omega'\cos\theta'\sin^2\theta'\Big[\sum_{n''=1}\,J_0\,\left[(C_2)(C_3)+(D_2)(D_3)\right]\nonumber\\
&&+\sum_{n''=1}\sum_{n=1}J_1{\cal B}\left[(C_2)_{n''}(D_3)_n+(C_3)_{n''}(D_2)_n\right]\Big]\Bigg\},
\label{dsdc}
\end{eqnarray}
where
\begin{eqnarray}
&&J_0\,=\,{(x'x)^{n''-1}\over[(n''-1)!]^2},
\nonumber\\
&&J_1\,=\,{(x'x)^{n''+n-1}\over(n''-1)!(n-1)!},
\nonumber\\
&&{\cal B}\,=\,\sum_{k=0}\,{(-x'x)^k\over k!(k+n''+n)!},
\end{eqnarray}
and the subscripts $n'',n$ in the double summations indicate which sum is used in $C_{1}\to C_{3}$ or $D_{1}\to D_{3}$, namely either $n''$ with $\epsilon''$ (as is) or $n''\to n$ with $\epsilon''\to\epsilon$ where $n,\,\epsilon$ are dummy variables.  For illustrative purposes, the upper limits on the sums for the $\perp'\perp$ mode are written down explicitly with the upper limit on the sum over $k$ in the Bessel function $\cal B$ being $n_{max}-n''-n+1$.  A value $n_{max}$ of 2 or 3 is sufficient for the range of $x',x$ chosen here, with an $n_{max}$ greater than this having no discernible effect on the differential cross sections.  The $\parallel'\parallel$ mode has terms with sums over $n''\geq0$ and these terms exhibit the PC divergence when $n''=0$ once $\omega'\geq2m/\sin\theta'$ is satisfied.  The differential cross section for this mode is
\begin{eqnarray}
&&{d\sigma^{\parallel'\parallel}\over d\cos\theta}={3\sigma_T\over4}\,{e^{-(x+x')}\omega^2\over32\,{\omega'}^2(\varepsilon'+m)}\nonumber\\
&&{1\over(m-\omega\sin^2\theta+\omega'-\omega'\cos\theta'\cos\theta)}\nonumber\\
&&\times\Bigg\{\cos^2\theta\cos^2\theta'\Big[\sum_{n''=1}\,J_0\,\left[(C_2)^2\,+\,(D_2)^2\right]\nonumber\\
&&-2\sum_{n''=1}\,\sum_{n=1}\,J_1\,{\cal B}\,(C_2)_{n''}(D_2)_n\Big]\nonumber\\
&&+\sin^2\theta\sin^2\theta'\Big[\sum_{n''=0}\,J_2\,\left[(C_1)^2\,+\,(D_1)^2\right]\nonumber\\
&&-2\sum_{n''=0}\,\sum_{n=0}\,J_3\,{\cal B}\,(C_1)_{n''}(D_1)_n\Big]\nonumber\\
&&+\,{\omega'}^2\cos^2\theta\sin^4\theta'\Big[\sum_{n''=1}\,J_0\,\left[(C_3)^2\,+\,(D_3)^2\right]\nonumber\\
&&-2\sum_{n''=1}\,\sum_{n=1}\,J_1\,{\cal B}\,(C_3)_{n''}(D_3)_n\Big]\nonumber\\
&&+\omega^2\sin^4\theta\cos^2\theta'\Big[\sum_{n''=1}\,J_0\,\left[(C_3)^2\,+\,(D_3)^2\right]\nonumber\\
&&-2\sum_{n''=1}\,\sum_{n=1}\,J_1\,{\cal B}\,(C_3)_{n''}(D_3)_n\Big]\nonumber\\
&&+2\omega'\omega\sin^2\theta'\sin^2\theta\cos\theta'\cos\theta\nonumber\\
&&\times\Bigg[-\sum_{n''=1}\,J_4\,\left[(C_1)(C_2)\,+\,(D_1)(D_2)\right]\nonumber\\
&&+\sum_{n''=0}\sum_{n=1}J_5{\cal B}\,\left[(C_1)_{n''}(D_2)_n+(D_1)_{n''}(C_2)_n\right]\nonumber\\
&&+\sum_{n''=1}J_0\left[(C_3)^2+(D_3)^2\right]\nonumber\\
&&-2\sum_{n''=1}\,\sum_{n=1}J_1{\cal B}\,(C_3)_{n''}(D_3)_n\Bigg]\nonumber\\
&&-2\omega\sin^2\theta\cos\theta\cos^2\theta'\nonumber\\
&&\times\Bigg[-\sum_{n''=1}J_0\left[(C_2)(C_3)+(D_2)(D_3)\right]\nonumber\\
&&+\sum_{n''=1}\sum_{n=1}J_1{\cal B}\left[(C_2)_{n''}(D_3)_n+(C_3)_{n''}(D_2)_n\right]\Bigg]
\nonumber\\
&&-2\omega'\sin^2\theta'\cos\theta'\cos^2\theta\nonumber\\
&&\times\Bigg[-\sum_{n''=1}J_0\left[(C_2)(C_3)+(D_2)(D_3)\right]\nonumber\\
&&+\sum_{n''=1}\sum_{n=1}J_1{\cal B}\left[(C_2)_{n''}(D_3)_n+(C_3)_{n''}(D_2)_n\right]\Bigg]
\nonumber\\
&&-2\omega\sin^2\theta\cos\theta\sin^2\theta'\nonumber\\
&&\times\Bigg[\sum_{n''=1}x'J_4\left[(C_1)(C_3)+(D_1)(D_3)\right]\nonumber\\
&&-\sum_{n''=0}\sum_{n=1}x'J_5{\cal B}\left[(C_1)_{n''}(D_3)_n+(D_1)_{n''}(C_3)_n\right]\Bigg]\nonumber\\
&&-2\omega'\sin^2\theta'\cos\theta'\sin^2\theta\nonumber\\
&&\times\Bigg[\sum_{n''=1}xJ_4\left[(C_1)(C_3)+(D_1)(D_3)\right]\nonumber\\
&&-\sum_{n''=0}\sum_{n=1}xJ_5{\cal B}\left[(C_1)_{n''}(D_3)_n+(D_1)_{n''}(C_3)_n\right]\Bigg]\Bigg\},
\label{parpar}
\end{eqnarray}
where
\begin{eqnarray}
&&J_2\,=\,{(x'x)^{n''}\over[n''!]^2},
\nonumber\\
&&J_3\,=\,{(x'x)^{n''+n}\over n''!\,n!},
\nonumber\\
&&J_4\,=\,{(x'x)^{n''-1}\over n''!\,(n''-1)!},
\nonumber\\
&&J_5\,=\,{(x'x)^{n''+n-1}\over n''!\,(n-1)!}.
\end{eqnarray}

Due to the restriction on $x'$ and $x$ in Eq.~(\ref{xxpineq}) and on $\omega'$ to be below the first GA resonance, only the first PC resonance at $n',\,n''=0$ is relevant in this study.  For the magnetic fields considered here, namely $0.1B_{cr},\,B_{cr},\,10B_{cr}$ and $100B_{cr}$, only the latter two exhibit the PC divergence. The PC divergence occurs at $\omega'_{PC}=2m/\sin\theta'$ when the intermediate particle is a positron in its ground state ($n''=0$).  The first PC threshold has no magnetic field dependence and only depends on the angle of the incident photon.  The only term that is resonant at $\omega'=\omega'_{PC}$ when $n'',\,n'=0$ and $\epsilon''=-$ is the first term in $[d^{+\,+}_{0\,0}]^{i\,j}_{n''}$, (Eq.~(\ref{d00})), which only contributes to the $\parallel'\,\parallel$ mode.  Hence the only mode affected by the lowest PC resonance is the $\parallel'\,\parallel$ mode.  The other modes only have PC resonances associated with either or both $n',\,n''>0$, which occurs at higher frequencies outside the regime considered here.  The only terms in Eq.~(\ref{parpar}) that  contribute to the 2-step process of pair creation and annihilation, in place of PC resonant Compton scattering once $\omega'\ge\omega'_{PC}$, are those involving $D_1$ when $n''$ or the dummy variable $n$ are zero and $\epsilon''=-$.  There are five such terms in Eq.~(\ref{parpar}).

\section{Resonant scattering as a 2-step process}
In a strong magnetic field, an electron in an excited state decays to its ground state on an extremely short timescale.  Hence, as is the usual procedure, the initial electron is assumed to be in its ground state ($n=0$) whereby its spin is antiparallel to the magnetic field ($\sigma=-1$).  Further, all evaluations are made in the rest frame of the initial electron so that its parallel momentum $p_z$ is zero.  Conservation of energy and parallel momentum between the initial and final states then dictates
\begin{eqnarray}
&&m\,+\,\omega'=\,\varepsilon'\,+\,\omega,\nonumber\\
&&\omega'\cos\theta'=\,p_z'\,+\,\omega\cos\theta,
\label{conserve}
\end{eqnarray}
where $\theta'$ and $\theta$ are the angles that the wave vectors of the initial and final photons make with the magnetic field in the rest frame of the initial electron.  There are two types of singularities that can occur in Compton scattering, associated with GA and PC thresholds, respectively.  These singularities appear as resonant denominators.  The lowest of the first type of singularity is dealt with by introducing a natural line width to the intermediate state; this line width is effectively the inverse of the lifetime of the intermediate state for gyromagnetic emission to a lower state.  The lowest of the second type of singularity has an intermediate state that is stable against gyromagnetic emission and has zero natural line width associated with gyromagnetic emission.  However a photon above the threshold for PC has a finite probability of decaying into a pair, with the pair subsequently annihilating into the final photon. The relevant lifetime in this case is that of the photon for decay into a pair.  The inverse of this lifetime leads to broadening effect, described by $\omega'\to\omega'-i\Gamma/2$.

\begin{figure}[t]
\begin{center}
\includegraphics[width=8cm]{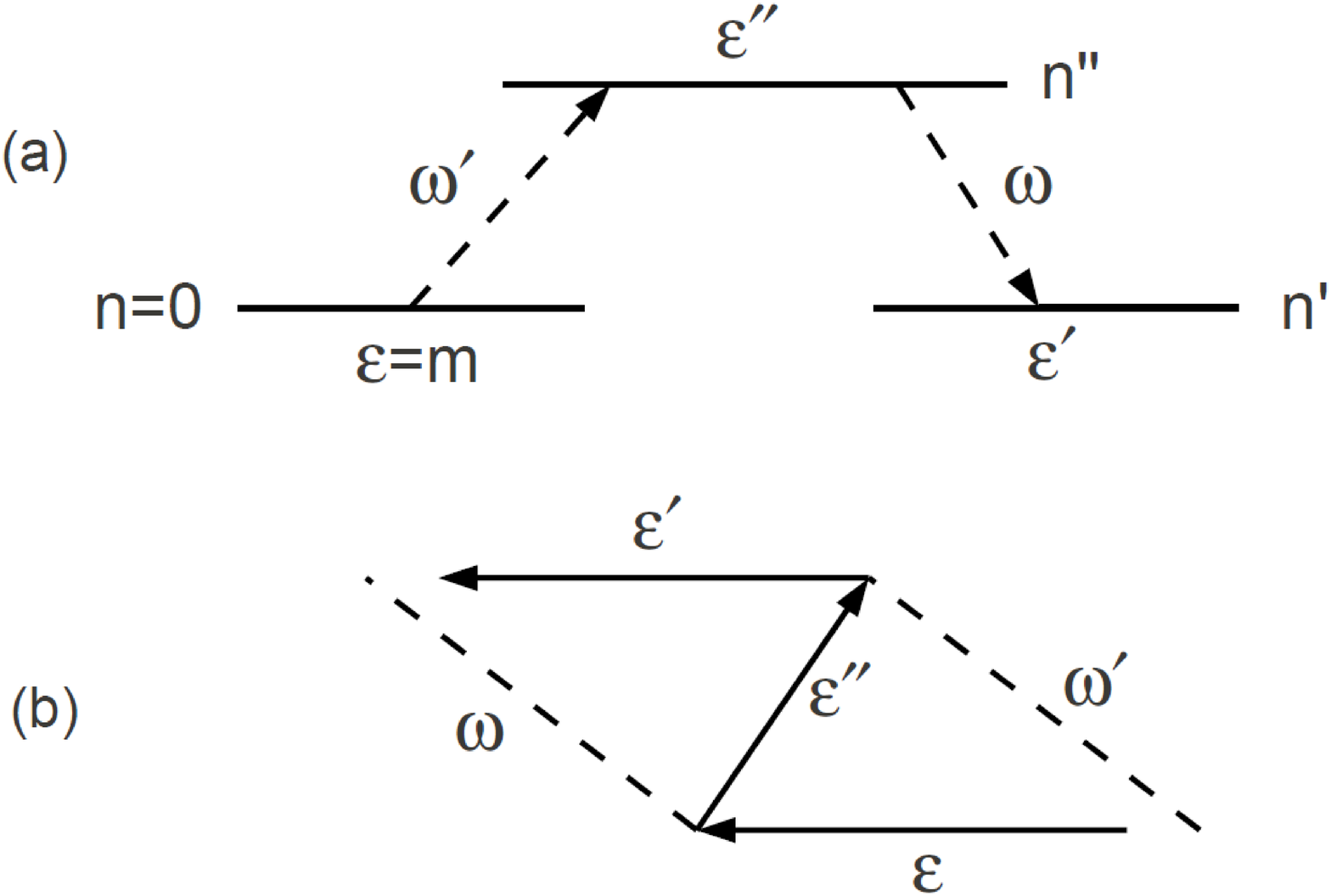}
\caption{(a) The 2-step process of gyromagnetic absorption and emission, and (b) the 2-step process of pair creation and annihilation.}
\label{fig:twostep}
\end{center}
\end{figure}

\subsection{Gyromagnetic (GA) resonance}
Consider the first Feynman diagram, Figure~\ref{fig:twostep}(a).  Should the intermediate particle be a real electron and its Landau quantum number be greater than the Landau quantum number of the initial and final electrons ($n''>n,\,n'$), then this first Feynman diagram can be cut at the intermediate electron state, so that it becomes two independent diagrams.  These two diagrams correspond to the 2-step process of gyromagnetic absorption of a photon of energy $\omega'$ followed by gyromagnetic emission of a photon of energy $\omega$.  The matrix element for the first Feynman diagram contains the denominator $\omega\,+\,\varepsilon'\,-\,\varepsilon''$, which is zero for the second step of this 2-step process, corresponding to the Compton scattering cross section being divergent.  The usual approach for dealing with this resonance is to introduce a natural line width to the intermediate state, so that the denominator is replaced according to 
\begin{equation}
\omega\,+\,\varepsilon'\,-\,\varepsilon''\,\to\,\omega\,+\,\varepsilon'\,-\,\varepsilon''\,+\,{i\Gamma/2},
\label{hfwd}
\end{equation}
with $\Gamma\,\to\,\Gamma_{\rm GE}$, where  $\Gamma_{\rm GE}$ is the gyromagnetic emission decay rate.  The process is regarded as Compton scattering for $|\omega\,+\,\varepsilon'\,-\,\varepsilon''|\gg\Gamma_{\rm GE}/2$ and as the 2-step process when this inequality is reversed.  A physical interpretation for this criterion is in terms of whether the intermediate state loses memory of how it was formed before it decays due to spontaneous emission.  One may interpret $\omega\,+\,\varepsilon'\,-\,\varepsilon''$ as a frequency mismatch that leads to loss of memory of how the intermediate state was formed, due to phase mixing. For $|\omega\,+\,\varepsilon'\,-\,\varepsilon''|\gg\Gamma_{\rm GE}/2$, the rate of decay due to spontaneous emission is negligible compared with the rate of phase mixing and the process must be regarded as scattering with a virtual intermediate state. When this inequality is reversed, the intermediate state decays faster than phase mixing occurs, and the `scattering' should then be regarded as a 2-step process.  Only incident photon frequencies below the first GA resonance at
\begin{equation}
\omega'_{GE}=\begin{cases}{\sqrt{m^2+2eB\sin^2\theta'}-m\over\sin^2\theta'}& \text{if $\theta'\neq 0^\circ,180^\circ$},\\
{eB\over m}& \text{if $\theta'=0^\circ,180^\circ$},\end{cases}
\end{equation}
are considered in the present paper, and the GA resonance is not relevant, but a PC resonance can occur. 

\subsection{First pair creation (PC) resonance}
Consider the second Feynman diagram, Figure~\ref{fig:twostep}(b), in which the intermediate particle is a positron. If the energy of the initial photon is greater than or equal to the PC threshold, this intermediate positron can be real rather than virtual. This corresponds to the second Feynman diagram being cut at the positron line so that it separates into two  diagrams, corresponding to the 2-step process of the creation of a pair by the initial photon and annihilation of a pair into the final photon. The matrix element for the second Feynman diagram includes an energy denominator $\varepsilon'+\varepsilon''-\omega'$, which is zero for the 2-step process, corresponding to the PC resonance.  The counterpart of the natural line width in this case is associated with the finite lifetime of the photon against decay into a pair, as the cut at the positron line involves three particles in their ground states. The broadening effect of PC is  included by giving the initial frequency an imaginary part, $\omega'\to\omega'-i\Gamma_{PC}/2$, where $\Gamma_{PC}$ is determined by the decay rate for pair creation.

When the line broadening effect is included, the resonant part of $[d^{+\,+}_{n'\,0}]^{i\,j}_{n''}$ with numerators of the general form $f+\varepsilon''g$ are modified as follows:
\begin{eqnarray}
&&{f+\varepsilon''g\over\varepsilon''(\varepsilon'+\varepsilon''-\omega')}\rightarrow{f(\varepsilon'+\varepsilon''-\omega')\over
{\varepsilon''}[(\varepsilon'+\varepsilon''-\omega')^2+\Gamma^2_{PC}/4]}\nonumber\\
&&\qquad\qquad\qquad+{g(\varepsilon'+\varepsilon''-\omega')\over[(\varepsilon'+\varepsilon''-\omega')^2+\Gamma^2_{PC}/4]}.
\end{eqnarray}
For $\Gamma_{PC}/2\ll|\varepsilon'+\varepsilon''-\omega'|$, the $\Gamma_{PC}/2$ correction may be ignored. For $\Gamma_{PC}/2\gg|\varepsilon'+\varepsilon''-\omega'|$ the PC resonance is treated in terms of a 2-step process.

There are notable differences between the GA and PC resonances.  Whereas $\Gamma_{GE}$ is independent of $\omega$ and $\theta$ and is calculable for $\omega'$ both below and above $\omega'_{GE}$, $\Gamma_{PC}$ depends on $\omega$ and $\theta$ and is only calculable once a threshold is reached.  Further, for the initial photon energies considered here, namely $\omega'<\omega'_{GE}$, the $n'=0$ and the $n''=0$ terms contribute only when the incident and final photons are both parallel-polarized.

\section{PC divergence}
Due to the restriction on $x'$ and $x$ in Eq.~(\ref{xxpineq}) and on $\omega'$ to be below the first GA resonance, only the first PC resonance at $n'=n''=0$ is relevant in this study.  For the magnetic fields considered here, namely $0.1B_{cr},\,B_{cr},\,10B_{cr}$ and $100B_{cr}$, the latter two exhibit the resonant Compton scattering associated with PC.  The term that has $n''=0$ when $n'=0$ is the first term in $[d^{+\,+}_{0\,0}]^{i\,j}_{n''}$ which contributes to the $\parallel'\,\parallel$ mode.  However, even when $\omega'$ is above $2m/\sin\theta'$, pair creation is possible only over a limited range of $\theta$.  From the conservation of energy equation for pair creation, $\varepsilon'+\varepsilon''=\omega'$, one obtains a $k_z-$dependent pair creation threshold value of $2(\varepsilon''-k_z\cos\theta')/\sin^2\theta'$ for the initial photon.  Since $\varepsilon'+\varepsilon''=\omega'$ is equivalent to $m+\varepsilon''=\omega$ from Eq.~(\ref{conserve}), one obtains the pair creation threshold of $\omega_{PC}=2m/\sin^2\theta$ for the final photon.  Equating this latter threshold of $\omega_{PC}$ to the right hand side of Eq.~(\ref{wfromwp}), yields a quadratic equation in $\cos\theta$ which allows one to evaluate the $\theta$-values at the PC thresholds, viz.
\begin{equation}
\cos\theta={2m\cos\theta'\pm\sqrt{{\omega'}^2\sin^4\theta'-4m^2\sin^2\theta'}\over(\omega'\sin^2\theta'+2m)}.
\label{costh12}
\end{equation}
The solutions are real if the quantity inside the square root is positive, which is satisfied when $\omega'\geq\omega'_{PC}=2m/\sin\theta'$.  When $\omega'=\omega'_{PC}$, Eq.~(\ref{costh12}) yields the one resonant solution for $\theta$.  When $\omega'>\omega'_{PC}$, the two resonant solutions for $\theta$, denoted by $\theta_1$ and $\theta_2$, yield a range of $\theta$, $\theta_1\le\theta\le\theta_2$, over which the 2-step process of pair creation and annihilation can compete with Compton scattering.  For $\theta$-values outside this range, one has $\omega<\omega_{PC}$ and pair creation is forbidden.  The $\parallel'\,\parallel$ differential cross sections for $\theta<\theta_1$ and $\theta>\theta_2$ become large as the angles $\theta=\theta_{1,2}$ are approached.

\begin{figure}[t]
\begin{center}
\includegraphics[width=8cm]{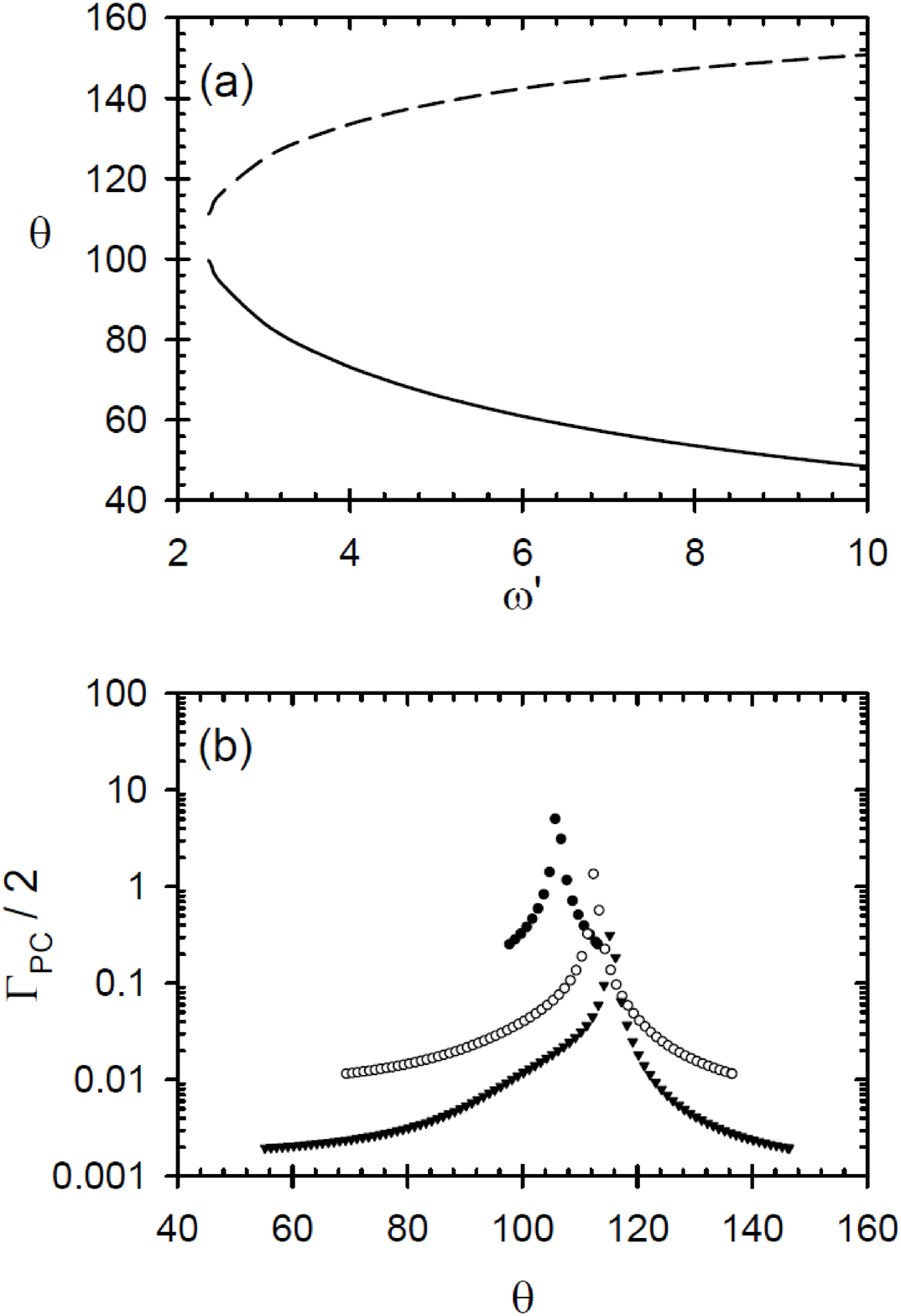}
\caption{(a) The PC threshold values of $\theta_1$ (solid line) and $\theta_2$ (dashed line) as a function of the incident photon energy $\omega'$ for $B=100B_{cr}$, $\theta'=120^\circ$.  As $\omega'$ increases so too does the range of $\theta$ over which the PC resonance occurs.  (b) $\Gamma_{PC}/2$ as a function of $\theta$ for $B=100B_{cr}$, $\theta'=120^\circ$ and the three $\omega'$ values (in units of $m$) above $\omega'_{PC}$ of $2.4m$ (closed circles), $4.5m$ (open circles) and $7.5m$ (closed triangles).  The magnitude of $\Gamma_{PC}/2$ decreases with increasing $\omega'$, and the range in $\theta$, over which the PC resonance can occur, increases.}
\label{fig:th1th2}
\end{center}
\end{figure}

In Figure~\ref{fig:th1th2}(a), the range of $\theta$ over which pair creation occurs as a function of $\omega'$ is plotted for the case when $B=100B_{cr}$ and $\theta'=120^\circ$.  As $\omega'$ increases so too does the range of $\theta$, $\theta_1\le\theta\le\theta_2$, over which PC can compete with Compton scattering.  As $\omega'$ becomes large, $\theta_1$ and $\theta_2$ approach $0^\circ$ and $180^\circ$ respectively.  The value of the PC decay rate varies over this range of $\theta$ according to Eq.~(\ref{Rat00}), having the same value at $\theta_1$ and $\theta_2$ and tending to infinity as the denominator in Eq.~(\ref{Rat00}) tends to zero.  This occurs at a $\theta$ value located between $\theta_1$ and $\theta_2$ when $\omega\cos\theta=k_z'\pm mk_z'/k_\perp'$, the sign of which depends upon whether $\theta'$ is below $\theta_1$, between $\theta_1$ and $\theta_2$ or above $\theta_2$.  The value, $\theta=\theta_0$, corresponding to $\Gamma_{PC}\to\infty$, follows from Eq.~(\ref{wfromwp}), and is given by
\begin{equation}
\cos\theta_0={u(\omega'+m\pm\sqrt{(\omega'\cos\theta'-u)^2+m^2}\,)\over({k'}_\perp^2+2m\omega'+2uk'_z-u^2)},
\end{equation}
with $u=\omega\cos\theta_0$ and where the relevant solution satisfies $\theta_1<\theta_0<\theta_2$.  In Figure~\ref{fig:th1th2}(b), the variation of $\Gamma_{PC}/2$ as a function of $\theta$, $\theta_1\le\theta\le\theta_2$, is presented for $B=100B_{cr}$, $\theta'=120^\circ$ and the three $\omega'$ values above $\omega'_{PC}$ of $2.4m$, $4.5m$ and $7.5m$.  The singular values occur at $\theta_0=106.1^\circ$, $112.6^\circ$ and $115.6^\circ$ respectively.

\begin{figure}[t]
\begin{center}
\includegraphics[width=8cm]{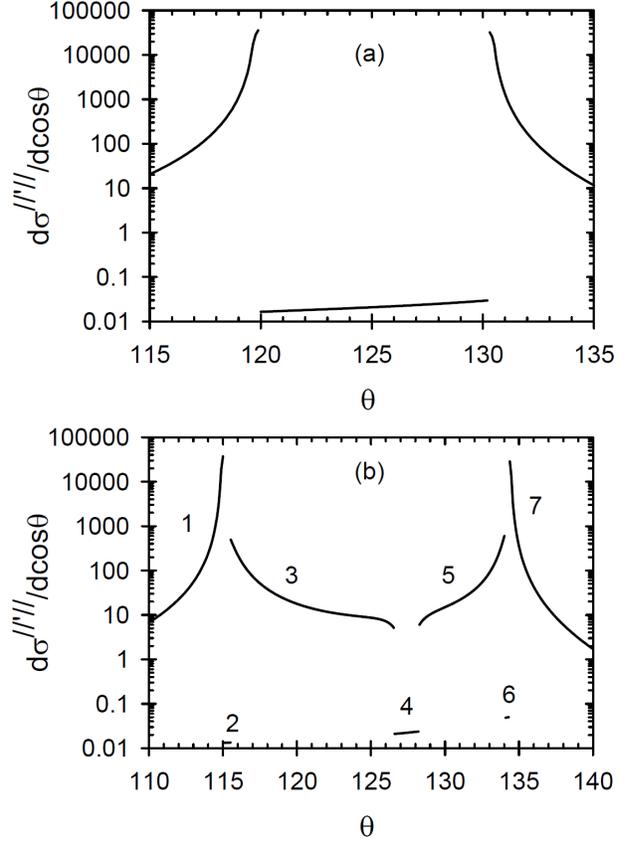}
\caption{Differential cross sections for the $\parallel'\,\parallel$ mode, $d\sigma^{\parallel'\parallel}/d\cos\theta$, as a function of $\theta$ for $B=10B_{cr}$, $\theta'=150^\circ$ and $\omega'$ values of (a) $4.1m$ and (b) $4.35m$.  As the PC thresholds are approached, the differential cross sections increase.  Between $\theta_1$ and $\theta_2$ where pair creation is allowed, there are regions where the 2-step process of pair creation and annihilation occurs and no resonant Compton scattering takes place, namely $\theta_1<\theta<\theta_2$ in (a) and regions 2, 4 and 6 in (b).}
\label{fig:dsth1th2}
\end{center}
\end{figure}

In Figures~\ref{fig:dsth1th2}(a) and (b), the differential cross section for the $\parallel'\parallel$ mode, $d\sigma^{\parallel'\parallel}/d\cos\theta$, is plotted in the region from just below $\theta_1$ to just above $\theta_2$ for $B=10B_{cr}$, $\theta'=150^\circ$ and the two values of $\omega'$, $4.10m$ and $4.35m$, respectively.  The pair creation contribution between $\theta_1$ and $\theta_2$ is treated either as a 2-step process (for $\Gamma_{PC}/2>|\varepsilon'+\varepsilon''-\omega'|$) or as a line width adjustment of the energy denominator (for $\Gamma_{PC}/2<|\varepsilon'+\varepsilon''-\omega'|$).  For $|\varepsilon'+\varepsilon''-\omega'|>10\Gamma_{PC}/2$, the $\Gamma_{PC}/2$ correction has little effect on the terms and the denominator is unadjusted.

For $\omega'=4.10m$, there is no scattering contribution from the $n'=n''=0$ term associated with the PC resonance as $\Gamma_{PC}/2>|\varepsilon'+\varepsilon''-\omega'|$ between $\theta_1$ and $\theta_2$, as shown in Figure~\ref{fig:dsth1th2}(a).  As the value of $\Gamma_{PC}/2$ decreases with increasing frequency $\omega'$, one expects Compton scattering to be more competitive with the 2-step process for $\omega'=4.35m$ than for $\omega'=4.1m$.  This is clearly evident in Figure~\ref{fig:dsth1th2}(b) for $\omega'=4.35m$, where there are two regions between $\theta_1$ and $\theta_2$ in which Compton scattering competes with the 2-step process.  These regions are denoted by 3 and 5.  In regions 2, 4 and 6, the PC resonance does not contribute to Compton scattering; only the 2-step process occurs.  Region 4 is in the vicinity of $\theta_0$ where $\Gamma_{PC}\to\infty$.  Regions 1 and 7 are outside the interval $\theta_1\le\theta\le\theta_2$, and only Compton scattering applies.  When one evaluates the $\parallel'\,\parallel$ contribution to the total cross section $\sigma/\sigma_T$ for $\omega'>\omega'_{PC}$, the areas of the different disjoint regions need to be evaluated separately and then added together.

As $\theta\to\theta_1$ from below and $\theta\to\theta_2$ from above, the differential cross sections for the $\parallel'\parallel$ mode tend to infinity as the PC resonances are approached.  One consistent means of dealing with these divergencies is as follows:

\noindent (i) using $\theta_1$ or $\theta_2$, determine $\Gamma_{PC}/2$ using Eq.~(\ref{Rat00}); the $\Gamma_{PC}/2$ values at $\theta_1$ and $\theta_2$ being identical.

\noindent (ii) determine $\omega_1|_{HM}$ and $\omega_2|_{HM}$, these being half the maximum values (HM) at $\omega_1$ and $\omega_2$ respectively, via:

(a) if $\theta'<\theta_1$ such that $\theta'<\theta_1<\theta_2<\theta_2|_{HM}$ then 
$$\omega_2|_{HM}=\omega_2-\Gamma_{PC}/2,$$
$$\omega_1|_{HM}=\begin{cases}\omega_1+\Gamma_{PC}/2& \text{if $\omega'-\omega_1\ge\Gamma_{PC}/2$},\\
2\omega'-\Gamma_{PC}/2-\omega_1& \text{if $\omega'-\omega_1<\Gamma_{PC}/2$},\end{cases}$$

(b) if $\theta'>\theta_2$ such that $\theta_1|_{HM}<\theta_1<\theta_2<\theta'$ then 
$$\omega_1|_{HM}=\omega_1-\Gamma_{PC}/2,$$
$$\omega_2|_{HM}=\begin{cases}\omega_2+\Gamma_{PC}/2& \text{if $\omega'-\omega_2\ge\Gamma_{PC}/2$},\\
2\omega'-\Gamma_{PC}/2-\omega_2& \text{if $\omega'-\omega_2<\Gamma_{PC}/2$},\end{cases}$$

(c) if $\theta_1\leq\theta'\leq\theta_2$ then 
$$\omega_1|_{HM}=\omega_1-\Gamma_{PC}/2,$$
$$\omega_2|_{HM}=\omega_2-\Gamma_{PC}/2.$$

\noindent (iii) substitute these values of $\omega_1|_{HM}$ and $\omega_2|_{HM}$ as the left hand sides of Eq.~(\ref{wfromwp}) to solve for $\theta_1|_{HM}$ and $\theta_2|_{HM}$ via
$$\cos\theta|_{HM}={k_z'\pm\sqrt{(\omega'-A)(\omega'-A+2m)}\over A}$$
where $A$ represents either $\omega_1|_{HM}$ or $\omega_2|_{HM}$ with $\theta|_{HM}$, the appropriate $\theta$-value.  The choice of the sign $\pm$ is determined as follows

(a) if $\theta'<\theta_1$, then one has $\omega_2|_{HM}$ from the $-$ branch and $\omega_1|_{HM}$ from the $\pm$ branch if $\omega_1+\Gamma_{PC}/2\gl\omega',$

(b) if $\theta'>\theta_2$, then one has $\omega_1|_{HM}$ from the $+$ branch and $\omega_2|_{HM}$ from the $\mp$ branch if $\omega_2+\Gamma_{PC}/2\gl\omega',$

(c) if $\theta_1\leq\theta'\leq\theta_2$, then one has $\omega_1|_{HM}$ from the $+$ branch and $\omega_2|_{HM}$ from the $-$ branch.

\noindent (iv) evaluate $d\sigma^{\parallel'\parallel}/d\cos\theta$ at $\theta_1|_{HM}$ and $\theta_2|_{HM}$, the values of which are then doubled to give the $d\sigma^{\parallel'\parallel}/d\cos\theta$ values at the resonances at $\theta_1$ and $\theta_2$ respectively.

Clearly the 2-step process of PC occurs at $\theta_1$ and $\theta_2$ (the $|\varepsilon'+\varepsilon''-\omega'|$ values being zero there), so that the contributions to the differential cross sections at these $\theta$ values due to Compton scattering are strictly zero.  However, as $\theta_1$ and $\theta_2$ are approached from below and above respectively, these resonant differential cross sections tend to infinity.  When $\omega'$ is equal to $\omega'_{PC}$, there is only the one resonant value, the $\Gamma_{PC}$ value of which tends to infinity and the above procedure does not work.  Consequently no cross sections are evaluated at this singularity but rather at $\omega'$ values within $0.05m$ of $\omega'_{PC}$ for given $B$ and $\theta'$ values.

\section{Results}
In order to obtain the total cross section for Compton scattering $\sigma$, the integrals of the differential cross sections $d\sigma^{M'M}/d\cos\theta$ over $\cos\theta$ between $-1$ and $1$, viz.
$$\sigma^{M'M}=\int_{-1}^1\,d\cos\theta\ {d\sigma^{M'M}\over d\cos\theta},$$
are evaluated as areas under the $d\sigma^{M'M}/d\cos\theta$ curves.  Since the $d\sigma^{M'M}/d\cos\theta$ values are smoothly varying with $\cos\theta$ for $M'M=\perp'\perp,\perp'\parallel,\parallel'\perp$, they are evaluated in $0.1^\circ$ steps from $180^\circ$ to $0^\circ$.  When $\omega'$ is below the PC threshold of $2m/\sin\theta'$, the $d\sigma^{\parallel'\parallel}/d\cos\theta$ values are likewise smoothly varying and are evaluated in $0.1^\circ$ steps.  This is also true when $\omega'$ is above the PC threshold for those components of $d\sigma^{\parallel'\parallel}/d\cos\theta$ that do not involve $(D_1)_0$, $\epsilon''=-$ in Eq.~(\ref{parpar}).  Those components involving $(D_1)_0$, $\epsilon''=-$, five in total, are combined and treated as follows.  Firstly, the range of angles, $180^\circ\geq\theta\geq0^\circ$, of the final photon is broken up into five distinct regions, and the contributions from $(D_1)_0$ to the cross section, denoted by $d\sigma^{\parallel'\parallel}/d\cos\theta|_{(D_1)_0}$, are evaluated in each region.  The first region is $180^\circ\geq\theta>\theta_2|_{HM}$, for which the $d\sigma^{\parallel'\parallel}/d\cos\theta|_{(D_1)_0}$ values are evaluated in $0.1^\circ$ steps as they are smoothly varying.  The second region is $\theta_2|_{HM}\geq\theta\geq\theta_2$, with
\begin{equation}
{d\sigma^{\parallel'\parallel}\over d\cos\theta}\Bigg|_{(D_1)_0}(\theta_2)=2\ {d\sigma^{\parallel'\parallel}\over d\cos\theta}\Bigg|_{(D_1)_0}(\theta_2|_{HM}).
\label{dsLor}
\end{equation}
If $\Delta\theta_2\equiv\theta_2|_{HM}-\theta_2$ satisfies $\Delta\theta_2\geq0.2^\circ$, then, to improve the accuracy of the final total cross section evaluation, $d\sigma^{\parallel'\parallel}/d\cos\theta|_{(D_1)_0}$ is evaluated at intermediate angles between $\theta_2|_{HM}$ and $\theta_2$ in $\sim0.1^\circ$ steps.  These intermediate values are obtained via a Lorentzian curve which satisfies Eq.~(\ref{dsLor}) and is subsequently of the form
\begin{eqnarray}
&&{d\sigma^{\parallel'\parallel}\over d\cos\theta}\Bigg|_{(D_1)_0}(\theta)={1\over2}\ {\Gamma^2_{PC}\over(\omega-\omega_2)^2+(\Gamma_{PC}/2)^2}\nonumber\\ 
&&\qquad\qquad\qquad\qquad\times{d\sigma^{\parallel'\parallel}\over d\cos\theta}\Bigg|_{(D_1)_0}(\theta_2|_{HM}),
\label{Lor}
\end{eqnarray}
with $\omega$ given by Eq.~(\ref{wfromwp}) for each $\theta$.  Should $\theta'$ be located between $\theta_2$ and $\theta$, then $(\omega-\omega_2)$ in Eq.~(\ref{Lor}) is replaced by $(2\omega'-\omega-\omega_2)$.  The third region, $\theta_2>\theta>\theta_1$, is where the 2-step PC process competes with Compton scattering and may comprise a number of disjoint contributions.  It is broken up into $0.01^\circ$ steps.  The fourth region, $\theta_1\geq\theta\geq\theta_1|_{HM}$, is treated in a similar way as the second region.  The fifth and final region, $\theta_1|_{HM}>\theta\geq0^\circ$, is treated in the same way as the first region.  

\begin{figure}[t]
\begin{center}
\includegraphics[width=8cm]{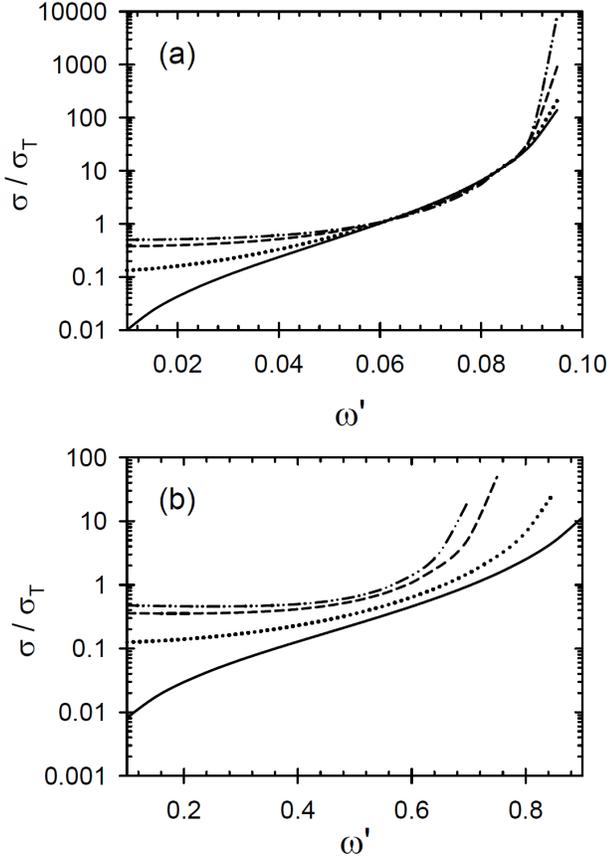}
\caption{Total Compton scattering cross section (in units of the Thomson cross section $\sigma_T$) as a function of the incident photon energy $\omega'$ (in units of $m$) for the magnetic field values of (a) $0.1B_{cr}$ and (b) $B_{cr}$, and a range of incident photon angles $\theta'$ comprising $\theta'=180^\circ,\,0^\circ$ (solid lines), $150^\circ,\,30^\circ$ (dotted lines), $120^\circ,60^\circ$ (dashed lines) and $90^\circ$ (dashed-dotted lines).}
\label{fig:Bpt11ptcs}
\end{center}
\end{figure}

\begin{figure}[t]
\begin{center}
\includegraphics[width=8cm]{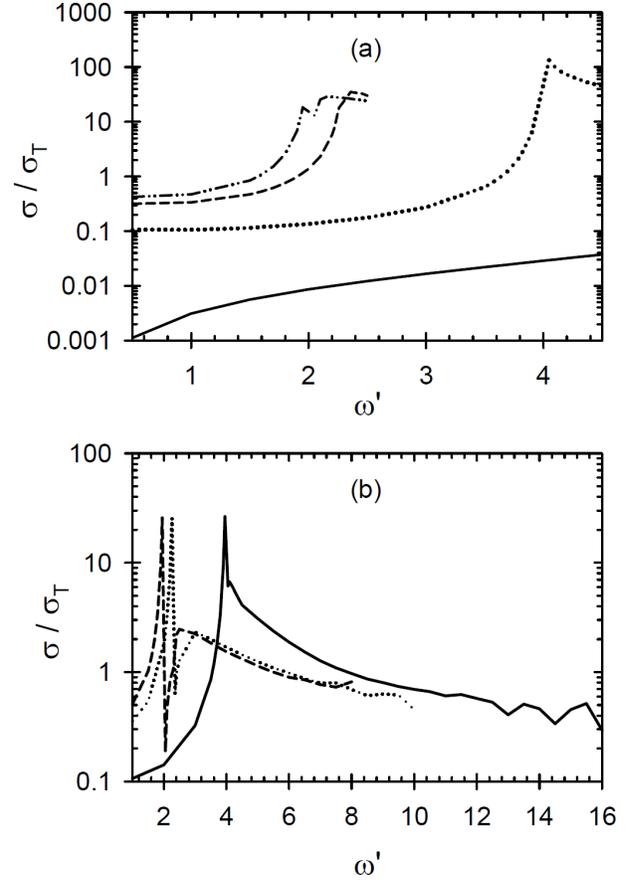}
\caption{Total Compton scattering cross section (in units of the Thomson cross section $\sigma_T$) as a function of the incident photon energy $\omega'$ (in units of $m$) for the magnetic field values of (a) $10B_{cr}$ and (b) $100B_{cr}$, and a range of incident photon angles $\theta'$ comprising $\theta'=180^\circ,\,0^\circ$ (solid line in (a)), $150^\circ,\,30^\circ$ (dotted line in (a) and solid line in (b)), $120^\circ,60^\circ$ (dashed line in (a) and dotted line in (b)) and $90^\circ$ (dashed-dotted line in  (a) and dashed line in (b)).}
\label{fig:B10pt100ptcs}
\end{center}
\end{figure}

The total cross section for photons with frequencies below the GA resonance ($n'=0,\ \sigma'=-$) is the average sum
$$\sigma\,=\,{1\over2}\sum_{M'M}\,\sigma^{M'M}.$$
The results of such an evaluation, expressed in units of the Thomson scattering cross section $\sigma_T$, are presented in Figure~\ref{fig:Bpt11ptcs} and Figure~\ref{fig:B10pt100ptcs} for $B/B_{cr}=0.1,1$ and $B/B_{cr}=10,100$ respectively for a range of $\omega'$ values below the first gyromagnetic resonance that satisfy $x'\la0.3$.  These cross sections are evaluated for four $\theta'$ values, namely $\theta'=180^\circ,\,0^\circ$ (solid lines), $150^\circ,\,30^\circ$ (dotted lines), $120^\circ,60^\circ$ (dashed lines) and $90^\circ$ (dashed-dotted lines) in Figures~\ref{fig:Bpt11ptcs} and Figure~\ref{fig:B10pt100ptcs}(a), and $\theta'=150^\circ,\,30^\circ$ (solid lines), $120^\circ,\,60^\circ$ (dotted lines) and $90^\circ$ (dashed lines) in Figure~\ref{fig:B10pt100ptcs}(b).  The results for $\theta'=180^\circ,\,0^\circ$ are omitted in Figure~\ref{fig:B10pt100ptcs}(b), being less than $0.001$ in magnitude.  The total cross sections are symmetric about $\theta'=90^\circ$, and only $180^\circ\geq\theta'\geq90^\circ$ is discussed explicitly.  The total cross sections increase in strength as $\theta'$ decreases from $180^\circ$ to $90^\circ$, that is, as the initial photon becomes increasingly oblique.

The results for $B=0.1B_{cr}$ and $B=B_{cr}$ are shown in Figures~\ref{fig:Bpt11ptcs}(a) and (b) respectively.  The cross sections become large as the GA resonance is approached.  The resonant frequency, $\omega'_{GE}$, decreases as $\theta'$ becomes more oblique.  The PC resonance at $\omega'_{PC}=2m/\sin\theta'$ is above the GA resonance values for $B/B_{cr}=0.1,1.0$, and is not shown.  There is also enhancement for oblique initial photon angles in the energy region $0<\omega'\la\omega'_{GE}/2$.  At the lowest frequencies, Figure~\ref{fig:Bpt11ptcs} compares favourably with the results of \citet{H79} and \citet{DH86}.  When $\theta'=180^\circ$, one has $x'=0$ and there is no restriction on the energy of the incident photon.  Only the $\theta'=180^\circ$ results for $\omega'<\omega'_{GE}$ are presented in Figures~\ref{fig:Bpt11ptcs}.  The cross sections over the extended range of incident photon energies, are presented as the dotted lines in Figures~\ref{fig:GonthlowB}.

The results for the two highest magnetic fields, $B=10B_{cr},100B_{cr}$ are shown in Figures~\ref{fig:B10pt100ptcs}.  For oblique incident photon angles, the restriction $x'\la0.3$ limits the energies of the incident photons to well below the GA resonances, which are not shown.  (This restriction does not apply when $\theta'=180^\circ$ as shown by the dotted lines in Figure~\ref{fig:GonthhiB}.)  The PC resonances are within the range of energies considered in Figure~\ref{fig:B10pt100ptcs} and are particularly apparent in Figure~\ref{fig:B10pt100ptcs}(b).  As these PC thresholds are approached, the cross sections are enhanced by at least 2 to 3 and 3 to 4 orders of magnitude over the $\theta'=180^\circ$ results when $B=10B_{cr}$ and $B=100B_{cr}$ respectively (see Figure~\ref{fig:B10pt100ptcs}).  These PC resonances result from the $(D_1)_{n''}$ or $(D_1)_n$ terms in the $\parallel'\parallel$ mode contribution given by Eq.~(\ref{parpar}) when $n''$ or the dummy variable $n$ are equal to zero and $\epsilon''=-$.

Once $\omega'$ is above the PC threshold and $\theta$ is between $\theta_1$ and $\theta_2$, the 2-step process of pair creation and annihilation competes with the Compton scattering process.  It is this competition that results in the decrease in the total Compton scattering cross section above the PC threshold.  For $B=10B_{cr}$, the decline in strength is gradual.  For $B=100B_{cr}$, the decrease is initially very steep but then settles down to a gradual decrease (see Figure~\ref{fig:B10pt100ptcs}(b)) with the cross section values at the endpoints (which satisfy $x'\la0.3$) similar in value to those of the lowest frequencies.  There is some variation in $\sigma/\sigma_T$ at the high frequency end for $B=100B_{cr}$, particularly evident when $\theta=150^\circ$.  There are two factors that may cause this variation: the extent of the PC competition region ($\theta_1\leq\theta\leq\theta_2$) which increases for a given $\theta'$ as $\omega'$ increases; and, the magnitude of $\Gamma_{PC}/2$ which decreases as $\omega'$ increases.  These effects are shown in Figure~\ref{fig:th1th2}.

The magnitudes of the total cross sections at the lowest frequencies for $\theta'=180^\circ$ decrease significantly as the strength of the magnetic field increases.  This is in contrast to what occurs at the lowest frequencies for the oblique angles considered here, where there are only small variations in the magnitudes of the cross sections as the strength of the magnetic field is increased.

\begin{figure}[t]
\begin{center}
\includegraphics[width=8cm]{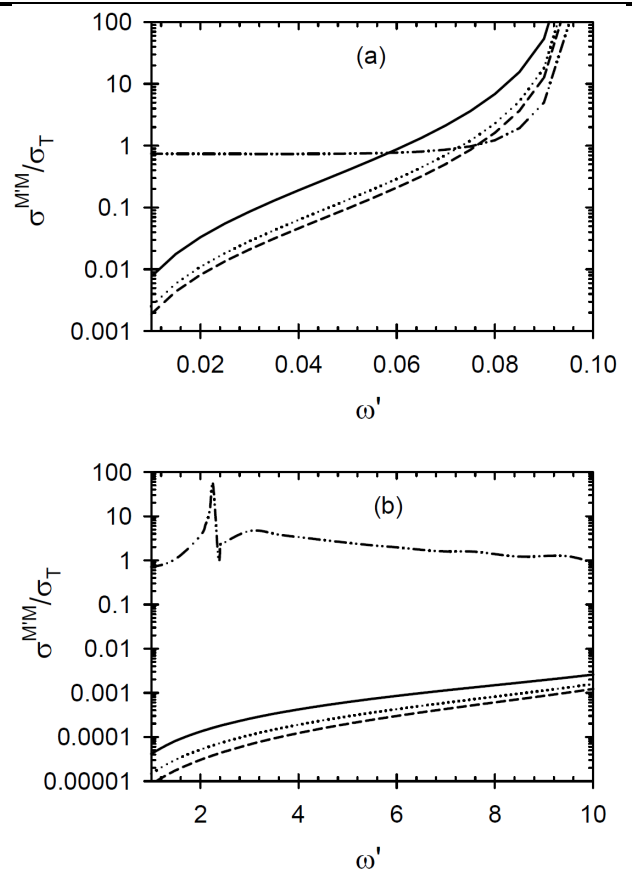}
\caption{Partial Compton scattering cross sections (in units of the Thomson cross section $\sigma_T$) as functions of the incident photon energy $\omega'$ (in units of $m$) for the modes $\perp'\,\perp$ (solid line), $\perp\,\parallel$ (dotted line), $\parallel'\,\perp$ (dashed line) and $\parallel'\,\parallel$ (dashed-double-dotted line), $\theta'=120^\circ$ and the two magnetic field values of (a) $B=0.1B_{cr}$ and (b) $B=100B_{cr}$.}
\label{fig:th120}
\end{center}
\end{figure}

The total cross section, $\sigma^{M'M}/\sigma_T$, for each polarization mode are found by integrating the differential cross sections over $\cos\theta$ numerically.  They vary as a function of the incident photon frequency $\omega'$ below the GA resonance, incident photon angle $\theta'$ and magnetic field $B$. The general trends are as follows.  Firstly when $\theta'=180^\circ$, the cross sections for the $\perp'\perp$ and $\parallel'\perp$ modes are identical, as are the $\perp'\parallel$ and $\parallel'\parallel$ modes.  As the strength of the magnetic field increases by an order of magnitude, the strengths of these cross sections decrease by an order of magnitude.  At the lowest frequencies considered for oblique incidence, the cross section for the $\parallel'\parallel$ mode is up to 2 orders of magnitude larger than the other three cross sections, which all increase with $\omega'$.  The $\perp'\perp$ mode overtakes the $\parallel'\parallel$ mode at $\omega'\approx0.05m,\,0.5m$ for $B=0.1B_{cr},\,B_{cr}$, respectively, as shown Figure~\ref{fig:th120}(a) for the case $B=0.1B_{cr}$, $\theta'=120^\circ$ and $\omega'<0.1m$.  The contributions from the $\parallel'\parallel$ mode thereafter closely follow the other modes as $\omega'_{GE}$ is approached.  At the two higher magnetic field values $B=10B_{cr},100B_{cr}$, $\omega'_{GE}$ is above the range of interest, and also above the first PC threshold at $\omega'_{PC}=2m/\sin\theta'$, which is in the range of interest.  The relative contributions from the three modes $\perp'\perp,\,\perp'\parallel$ and $\parallel'\perp$ is similar to the cases $B=0.1B_{cr},B_{cr}$, with the $\perp'\perp$ mode having the largest cross section.  The contribution from the $\parallel'\parallel$ mode is enhanced by 1 to 2 orders of magnitude for $=10B_{cr},100B_{cr}$ at the lowest $\omega'$ considered and is largely independent of the field strength.  This contribution continues to increase to a peak at $\omega'_{PC}$ and remains larger than the other three contributions for $\omega'>\omega'_{PC}$. These features are illustrated in Figure~\ref{fig:th120}(b) for the case $B=100B_{cr}$, $\theta'=120^\circ$ and $\omega'\la10m$.  Clearly, the $\sigma/\sigma_T$ results at oblique incident angles are dominated by the $\parallel'\parallel$ contribution.

\begin{figure}[t]
\begin{center}
\includegraphics[width=8cm]{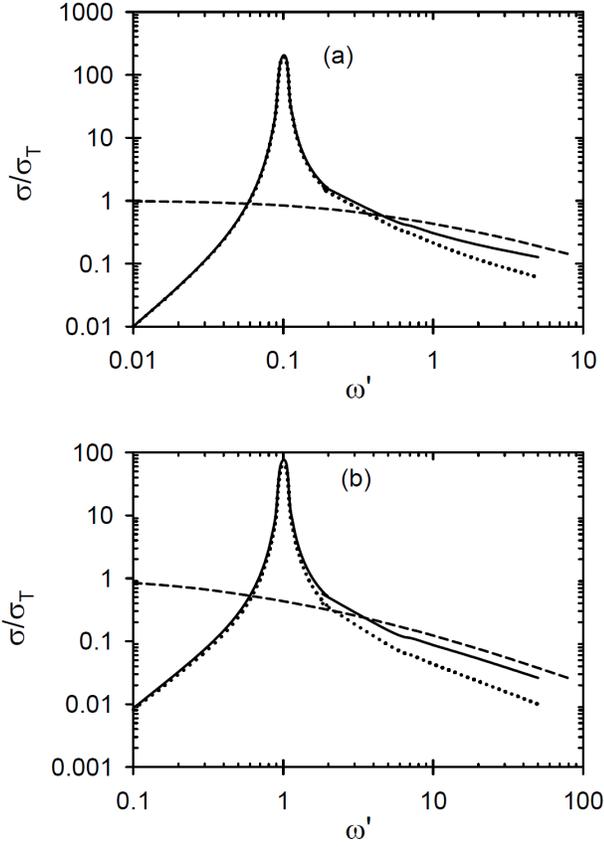}
\caption{Comparison of the total Compton scattering cross sections as a functin of $\omega'$ evaluated using either the factor $\omega/\omega'$ (solid lines) (\cite{G00}) or the factor $\omega^2/{\omega'}^2$ (dotted lines) for $\theta'=180^\circ,\,0^\circ$ and (a) $B=0.1B_{cr}$, (b) $B=B_{cr}$.  The Klein-Nishina cross sections (dashed lines) are included for comparison.}
\label{fig:GonthlowB}
\end{center}
\end{figure}

\begin{figure}[t]
\begin{center}
\includegraphics[width=8cm]{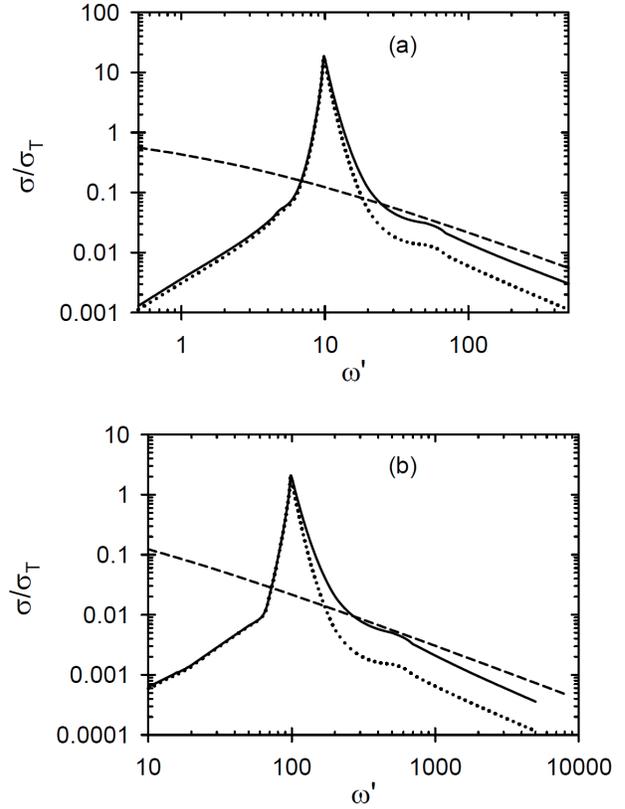}
\caption{Comparison of the total Compton scattering cross sections as a functin of $\omega'$ evaluated using either the factor $\omega/\omega'$ (solid lines) (\cite{G00}) or the factor $\omega^2/{\omega'}^2$ (dotted lines) for $\theta'=180^\circ,\,0^\circ$ and (a) $B=10B_{cr}$, (b) $B=100B_{cr}$.  The Klein-Nishina cross sections (dashed lines) are included for comparison.}
\label{fig:GonthhiB}
\end{center}
\end{figure}

In comparing the foregoing results with those of others, one needs to comment further on the results of \citet{G00}. Comparison is complicated by their use of an incorrect factor, $\omega/\omega'$ instead of the correct factor $\omega^2/{\omega'}^2$, in  their cross sections, which are for the special case $\theta'=180^\circ$ in the notation used here.  In Figures~\ref{fig:GonthlowB} and \ref{fig:GonthhiB}, the total Compton scattering cross sections $\sigma/\sigma_T$ are plotted as a function of the frequency of the incident photon $\omega'$ for $\theta'=180^\circ$ and a range of magnetic fields, $B/B_{cr}=0.1,1$ and $10,100$ respectively.  The solid and dotted lines are the cross sections evaluated using the factor $\omega/\omega'$ and $\omega^2/{\omega'}^2$ respectively.  The two results are similar for $\omega'<eB/m$.  For $\omega'>eB/m$,  the results using the factor $\omega/\omega'$ are enhanced significantly over the results using the factor $\omega^2/{\omega'}^2$.  It is interesting that the error introduced by using the incorrect factor lead to cross sections that are closer to the Klein-Nishina cross sections (see dashed lines) at the higher frequencies than are the correct results.  However, there is no reason why the results should be the same as the Klein-Nishina cross section as the latter does not include the effect of the magnetic field.  The formula for the Klein-Nishina cross section for unpolarized photons is given by
\begin{eqnarray}
&&\sigma_{KN}={3\sigma_T\over8}{m\over2\omega'}\Big\{2\left[{2m^2\over{\omega'}^2}+{2m\over\omega'}-1\right]\ln\left({m\over m+2\omega'}\right)
\nonumber\\
&&\qquad+1+{8m\over\omega'}-{m^2\over(m+2\omega')^2}\Big\}.
\end{eqnarray}
The value of $\sigma_{KN}$ approaches $\sigma_T$ as $\omega'\to0$ and it has no magnetic field dependence.  The effect of a magnetic field was acknowledged to be important by \citet{M74} who evaluated a magnetic field correction to the Klein-Nishina cross section albeit only for a weak field.

The results presented here have some overlap with those of \citet{CR09}, who evaluated photon absorptions for the different modes at a magnetic field of $200B_{cr}$ and $\theta'=90^\circ$.  They took into account wave function renormalisation and photon dispersion and included a plasma component.  At the lowest plasma temperature of $50keV$, the relative strengths of the contributions from the different modes follow the trends discussed above.  However,  renormalisation and dispersion, which are not included here, affect their Figures~3 and~4, making more detailed comparison difficult.

\section{Discussion}
The pair creation divergence present in Compton scattering was recognized by Herold in 1979 and it has been discussed by a number of authors (eg: \cite{NK93},\cite{GHS95},\cite{K96}).  The two most recent papers that attempted to tackle the divergence formally were the one by \citet{GHS95} and the other by \citet{K96}.  In the former paper, the authors abandoned energy conservation entirely and assigned a Weisskopf-Wigner decay width to the initial and final electrons and photons as well as the intermediate particle.  This yielded a scattering matrix element with a time dependence representing the time difference between the formation of the initial states and the measurement of the final states.  In the latter paper, Kachelriess introduced Licht fields and spectral functions.  The main point made in the present paper is that the PC resonance is analogous to the more familiar GA resonances in ``resonant Compton scattering.'' At the resonance, the scattering formally separates into a two-stage process, gyromagnetic absorption and emission at a GA resonance, and pair creation and annihilation at a PC resonance. The usual procedure of including the broadening of the resonance due to the finite lifetime of the intermediate state applies to both resonances, with the relevant lifetime at a PC resonance involving that due to pair creation.

In summary, due to the restriction imposed on $x',x$ and $n,\,n'$, only $\omega'$ values below the GA resonance are considered in this study.  Nevertheless, some important results are obtained when $\theta'$ is not aligned along the magnetic field.  In the energy regions where the PC resonance does not occur, the Compton scattering cross sections are larger by one to two orders of magnitude at the lower frequencies, irrespective of the strength of the magnetic field.  There is additional strength at the two highest magnetic fields at frequencies in the vicinity of the PC resonance.  Unlike the GA resonance which is approximately symmetric and of a finite width affecting all 4 possible polarization mode combinations, the lowest PC resonance has a strong broad tail and affects only the $\parallel'\,\parallel$ mode.  (Higher PC resonances affect the other modes.)  The PC divergence however is only apparent when the incident photon is not propagating parallel to the magnetic field.  The photon's angle of incidence need only be slightly oblique for the resonance associated with pair creation to dominate the Compton scattering cross section.  Under these conditions, Compton scattering becomes an important process for the energy conversion of parallel-polarized photons.  The PC resonance in Compton scattering needs to be taken into account when considering the propagation of high energy ($\gg$ few MeV) photons in superstrong magnetic fields ($B\gg B_{cr}$) in magnetars.

\section*{\it Acknowledgment}
The author thanks Professor Donald Melrose and Dr.~Qinghuan Luo for constructive comments on the manuscript.  Professor Melrose was always available for useful discussions as the work progressed.

\appendix
\section{$S$-matrix}
\label{A-appendix}
The $S$-matrix of order $n$ is of the form (\cite{QPDI})
\begin{eqnarray}
&&{\hat S}^{(n)}={(-ie)^n\over n!}\int d^4x_1\int d^4x_2\dots\int d^4x_n\nonumber\\
&&\qquad\times{\hat{\cal T}}\{:{\bar \psi}(x_1)\Az(x_1)\psi(x_1)::{\bar \psi}(x_2)\Az(x_2)\psi(x_2):\,\dots\,:{\bar \psi}(x_n)\Az(x_n)\psi(x_n):\},
\end{eqnarray}
where $\hat{\cal T}$ is the chronological operator.  The first order $S$-matrix element, with the one vertex, can be written as
\begin{equation}
S^{(1)}_{fi}=ie\int\,d^4x\,{\bar \psi}^{\epsilon'}_f(x)\,\Az(x)\,\psi^\epsilon_i(x)
\end{equation}
and, to lowest order, corresponds to 8 combinations (or Feynman diagrams): $\epsilon'=\pm$, $\epsilon=\pm$ and $\Az=\gamma^\mu A_\mu$ with $A^\mu(x)$ equal to either $\sqrt{\mu_0R_M({\bf k})/(\omega_MV)}e^\mu_M({\bf k})e^{-ik_Mx}$ for an incoming photon or $\sqrt{\mu_0R_M({\bf k})/(\omega_MV)}e^{\mu*}_M({\bf k})e^{ik_Mx}$ for an outgoing photon, where $R_M({\bf k})$ is the ratio of the electric to total energy of the mode and is approximated by one-half.  Each incoming electron or positron line with quantum number $q$ has an associated wave function $\psi^+_q({\bf x})e^{-i\varepsilon_qt}$ or ${\bar \psi}^-_q({\bf x})e^{-i\varepsilon_qt}$ respectively.  Each outgoing electron or positron line with quantum number $q'$ has an associated wave function ${\bar \psi}^+_{q'}({\bf x})e^{i\varepsilon'_{q'}t}$ or $\psi^-_{q'}({\bf x})e^{i\varepsilon'_{q'}t}$ respectively.  Two of these combinations correspond to gyromagnetic emission and pair creation.  The second order $S$-matrix element, with two vertices, of which Compton scattering is one case, can be written as
\begin{equation}
S^{(2)}_{fi}=(ie)^2\sum_{q'',\epsilon''}\int\int d^4x'\,d^4x\,[{\bar \psi}^{\epsilon'}(x')\,\Az(x')\,\underbrace{\psi^{\epsilon''}(x')]\,[{\bar \psi}^{\epsilon''}(x)}\,\Az'(x)\,\psi^\epsilon(x)]\,+\,\{A\leftrightarrow A'\},
\end{equation}
where the underbrace denotes a contraction.  For Compton scattering as described by the two Feynman diagrams in Figure~\ref{fig:Feyndiag}, the $S$-matrix element is made up of two terms, viz.
\begin{eqnarray}
&&S^{(2)}_{fi}=(ie)^2{\mu_0\over V}\sqrt{{R_M({\bf k})R_{M'}({\bf k}')\over\omega\omega'}}\sum_q\sum_{q'}\int\,d^4x'\,\int\,d^4x
\nonumber\\
&&\ \ \times\Bigg\{{\bar \psi}^{\epsilon'}_{q'}({\bf x'})e^{i\epsilon'\varepsilon'_{q'}t'}\gamma_\mu e^{*\mu}_M({\bf k})e^{i\omega t'-i{\bf k.x'}}iG(x',x)\gamma_\nu e^{\nu}_{M'}({\bf k}')e^{-i\omega't+i{\bf k'.x}}\psi^\epsilon_q({\bf x})e^{-i\epsilon\varepsilon_qt}\nonumber\\
&&\qquad+{\bar \psi}^{\epsilon'}_{q'}({\bf x'})e^{i\epsilon'\varepsilon'_{q'}t'}\gamma_\nu e^{\nu}_{M'}({\bf k}')e^{-i\omega't'+i{\bf k'.x'}}iG(x',x)\gamma_\mu e^{*\mu}_M({\bf k})e^{i\omega t-i{\bf k.x}}\psi^\epsilon_q({\bf x})e^{-i\epsilon\varepsilon_qt}\Bigg\},
\label{S2}
\end{eqnarray}
where the contraction $\underbrace{\psi^{\epsilon''}(x'){\bar \psi}^{\epsilon''}(x)}$ has been replaced by the electron propagator $iG(x',x)$ given by
\begin{equation}
G(x',x)=-i\sum_{q''}[\theta(t'-t)\psi^+_{q''}(x'){\bar \psi}^+_{q''}(x)e^{-i\varepsilon''(t'-t)}-\theta(t-t')\psi^-_{q''}(x'){\bar \psi}^-_{q''}(x)e^{i\varepsilon''(t'-t)}],
\end{equation}
for an electron ($\epsilon''=+$) or positron ($\epsilon''=-$) as the intermediate particle.  In Eq.~(\ref{S2}) and subsequent equations, the subscripts $M$ and $M'$ on the frequencies and wave vectors of the initial and final photons have been omitted for convenience.  The time-ordering step function $\theta(t)$ has the integral representation
\begin{equation}
\theta(t)=\lim_{\eta\to0,\eta>0}\,{i\over2\pi}\int_{-\infty}^\infty\,ds\,{e^{-ist}\over s+i\eta}.
\end{equation}
The time dependencies, $t$ and $t'$, are exponential and can be factorized out so that the integrals over $t$, $t'$ and $s$ can be evaluated as follows
\begin{eqnarray}
&&\qquad\int dt'\int dt\int ds\ e^{i\epsilon'\varepsilon't'}e^{i\omega t'}e^{-i\epsilon''\varepsilon''(t'-t)}e^{-i\omega't}e^{-i\epsilon\varepsilon t}{i\over2\pi}{e^{-i\epsilon''(t'-t)s}\over s+i\eta}\nonumber\\
&&\qquad\qquad={i\over2\pi}\int ds\,2\pi\delta(\epsilon'\varepsilon'+\omega-\epsilon''(\varepsilon''+s))\,2\pi\delta(\epsilon''\varepsilon''-\omega'-\epsilon\varepsilon+\epsilon''s){1\over s+i\eta}\nonumber\\
&&\qquad\qquad=2\pi i\,\delta(\epsilon'\varepsilon'+\omega-\epsilon\varepsilon-\omega'){\epsilon''\over\epsilon\varepsilon-\epsilon''\varepsilon''+\omega'+i\epsilon''\eta},
\end{eqnarray}
for the first term in Eq.~(\ref{S2}), and
\begin{eqnarray}
&&\qquad\int dt'\int dt\int ds\ e^{i\epsilon'\varepsilon't'}e^{-i\omega't'}e^{-i\epsilon''\varepsilon''(t'-t)}e^{i\omega t}e^{-i\epsilon\varepsilon t}{i\over2\pi}{e^{-i\epsilon''(t'-t)s}\over s+i\eta}\nonumber\\
&&\qquad\qquad={i\over2\pi}\int ds\,2\pi\delta(\epsilon'\varepsilon'-\omega'-\epsilon''(\varepsilon''+s))\,2\pi\delta(\epsilon''\varepsilon''+\omega-\epsilon\varepsilon+\epsilon''s){1\over s+i\eta}\nonumber\\
&&\qquad\qquad=2\pi i\,\delta(\epsilon'\varepsilon'-\omega'+\omega-\epsilon\varepsilon){\epsilon''\over\epsilon\varepsilon-\epsilon''\varepsilon''-\omega+i\epsilon''\eta},
\end{eqnarray}
for the second term.  The $\delta$-functions yield the conservation of energy equation for Compton scattering.  The $S$-matrix then becomes
\begin{eqnarray}
&&S^{(2)}_{fi}=(ie)^2\sum_q\sum_{q'}\sum_{q''}\sum_{\epsilon''}{\mu_0\over2V}\sqrt{{1\over\omega\omega'}}\int d{\bf x'}\int d{\bf x}\ 2\pi\delta(\epsilon'\varepsilon'+\omega-\epsilon\varepsilon-\omega')\nonumber\\
&&\times\Bigg\{{{\bar \psi}^{\epsilon'}_{q'}({\bf x'})\gamma_\mu e^{*\mu}_M({\bf k})e^{-i{\bf k.x'}}\psi^{\epsilon''}_{q''}({\bf x'}){\bar \psi}^{\epsilon''}_{q''}({\bf x})\gamma_\nu e^{\nu}_{M'}({\bf k}')e^{i{\bf k'.x}}\psi^\epsilon_q({\bf x})\over\epsilon\varepsilon-\epsilon''\varepsilon''+\omega'}\nonumber\\
&&\qquad\ \ +{{\bar \psi}^{\epsilon'}_{q'}({\bf x'})\gamma_\nu e^{\nu}_{M'}({\bf k}')e^{i{\bf k'.x'}}\psi^{\epsilon''}_{q''}({\bf x'}){\bar \psi}^{\epsilon''}_{q''}({\bf x})\gamma_\mu e^{*\mu}_M({\bf k})e^{-i{\bf k.x}}\psi^\epsilon_q({\bf x})\over\epsilon\varepsilon-\epsilon''\varepsilon''-\omega}\Bigg\}.
\label{S2a}
\end{eqnarray}

To carry out the spatial integrations, using the notation of \citet{MP83}, one defines
\begin{equation}
[\gamma^{\epsilon'\epsilon}_{q'q}(\pm{\bf k})]^\mu=\int d{\bf x}\ e^{\mp i{\bf k.x}}{\bar \psi}^{\epsilon'}_{q'}({\bf x})\gamma^\mu\psi^\epsilon_q({\bf x}),
\end{equation}
to obtain
\begin{equation}
[\gamma^{\epsilon'\epsilon}_{q'q}(\pm{\bf k})]^\mu={(2\pi)^2\over V\sqrt{eB}}e^{(\pm ik_x(\epsilon p_y+\epsilon'p'_y)/2eB)}\delta(\epsilon p_y-\epsilon'p'_y\mp k_y)\delta(\epsilon p_z-\epsilon'p'_z\mp k_z)[\Gamma^{\epsilon'\epsilon}_{q'q}(\pm{\bf k})]^\mu,
\end{equation}
where ${\bf k}=(k_\perp\cos\psi,k_\perp\sin\psi,k_z)$, $[\gamma^{\epsilon'\epsilon}_{q'q}(-{\bf k})]^\mu\equiv[\gamma^{\epsilon\epsilon'}_{qq'}({\bf k})]^{*\mu}$ and
\begin{eqnarray}
&&[\Gamma^{\epsilon'\epsilon}_{q'q}({\bf k})]^\mu=\left\{-ie^{-i\psi}\right\}^{n'-n}\nonumber\\
&&\qquad\qquad\qquad\times\big\{({C'}^*_1C_1+{C'}^*_3C_3)J^{n-1}_{n'-n}+({C'}^*_2C_2+{C'}^*_4C_4)J^n_{n'-n},\nonumber\\
&&\qquad\qquad\qquad i({C'}^*_1C_4+{C'}^*_3C_2)e^{i\psi}J^n_{n'-n-1}-i({C'}^*_2C_3+{C'}^*_4C_1)e^{-i\psi}J^{n-1}_{n'-n+1},\nonumber\\
&&\qquad\qquad\qquad({C'}^*_1C_4+{C'}^*_3C_2)e^{i\psi}J^n_{n'-n-1}+({C'}^*_2C_3+{C'}^*_4C_1)e^{-i\psi}J^{n-1}_{n'-n+1},\nonumber\\
&&\qquad\qquad\qquad({C'}^*_1C_3+{C'}^*_3C_1)J^{n-1}_{n'-n}-({C'}^*_2C_4+{C'}^*_4C_2)J^n_{n'-n}\big\}.
\label{Gammak}
\end{eqnarray}
The $J$-functions are defined as
\begin{equation}
J^\mu_\nu(x)=\sqrt{{\mu!\over(\mu+\nu)!}}\,\exp(-\half x)\,x^{\nu/2}\,L^\nu_\mu(x),
\label{Jfunct}
\end{equation}
with argument $x=k^2_\perp/2eB$, and $L^\nu_\mu(x)$ is the generalized Laguerre polynomial
\begin{equation}
L^\nu_\mu(x)=\sum_{m=0}^\mu\,
\left(
\begin{array}{c}
\mu+\nu\\
\mu-m
\end{array}
\right)
\,{(-x)^m\over m!}.
\end{equation}

The $S$-matrix element is now of the form
\begin{eqnarray}
&&S^{(2)}_{fi}=(ie)^2\sum_q\sum_{q'}\sum_{q''}\sum_{\epsilon''}{\mu_0\over2V}{1\over\sqrt{\omega'\omega}}\ 2\pi\delta(\epsilon'\varepsilon'+\omega-\epsilon\varepsilon-\omega'){1\over V^2eB}\nonumber\\
&&\times\Bigg\{{e^{*\mu}_M({\bf k})e^\nu_{M'}({\bf k}')\over\epsilon\varepsilon-\epsilon''\varepsilon''+\omega'}\,2\pi\delta(\epsilon''p''_y-\epsilon'p'_y-k_y)2\pi\delta(\epsilon''p''_z-\epsilon'p'_z-k_z)
e^{ik_x(\epsilon''p''_y+\epsilon'p'_y)/2eB}[\Gamma^{\epsilon'\epsilon''}_{q'q''}({\bf k})]^\mu\nonumber\\
&&\qquad\times2\pi\delta(\epsilon p_y-\epsilon''p''_y+k'_y)2\pi\delta(\epsilon p_z-\epsilon''p''_z+k'_z)e^{-ik'_x(\epsilon''p''_y+\epsilon p_y)/2eB}[\Gamma^{\epsilon\epsilon''}_{qq''}({\bf k}')]^{*\nu}\nonumber\\
&&+{e^{*\mu}_M({\bf k})e^\nu_{M'}({\bf k}')\over\epsilon\varepsilon-\epsilon''\varepsilon''-\omega}\,2\pi\delta(\epsilon''p''_y-\epsilon'p'_y+k'_y)2\pi\delta(\epsilon''p''_z-\epsilon'p'_z+k'_z)
e^{-ik'_x(\epsilon'p'_y+\epsilon''p''_y)/2eB}[\Gamma^{\epsilon''\epsilon'}_{q''q'}({\bf k}')]^{*\nu}\nonumber\\
&&\qquad\times2\pi\delta(\epsilon p_y-\epsilon''p''_y-k_y)2\pi\delta(\epsilon p_z-\epsilon''p''_z-k_z)e^{ik_x(\epsilon''p''_y+\epsilon p_y)/2eB}[\Gamma^{\epsilon''\epsilon}_{q''q}({\bf k})]^\mu\Bigg\}.
\end{eqnarray}
The sum over the intermediate states (\cite{MP83}) is
\begin{equation}
\sum_{q''}=\sum_{\sigma''=\pm}\sum_{n''=0}^\infty\,V(eB)^{1/2}\int{dp''_y\over2\pi}\int{dp''_z\over2\pi}.
\end{equation}
These integrals give
\begin{eqnarray}
&&\qquad\int{dp''_z\over2\pi}2\pi\delta(\epsilon''p''_z-\epsilon'p'_z-k_z)2\pi\delta(\epsilon p_z-\epsilon''p''_z+k'_z)=2\pi\delta(\epsilon p_z-\epsilon'p'_z+k'_z-k_z),\nonumber\\
&&\qquad\int{dp''_y\over2\pi}2\pi\delta(\epsilon''p''_y-\epsilon'p'_y-k_y)2\pi\delta(\epsilon p_y-\epsilon''p''_y+k'_y)e^{ik_x(\epsilon''p''_y+\epsilon'p'_y)/2eB}\,e^{-ik'_x(\epsilon''p''_y+\epsilon p_y)/2eB}\nonumber\\
&&\qquad\qquad=2\pi\delta(\epsilon p_y+k'_y-\epsilon'p'_y-k_y)e^{ik_x(\epsilon p_y+k'_y+\epsilon'p'_y)/2eB}\,e^{-ik'_x(\epsilon'p'_y+k_y+\epsilon p_y)/2eB},\nonumber\\
&&\qquad\qquad=2\pi\delta(\epsilon p_y+k'_y-\epsilon'p'_y-k_y)e^{i(k_x-k'_x)(\epsilon p_y+\epsilon'p'_y)/2eB}\,e^{i(k_xk'_y-k'_xk_y)/2eB},
\end{eqnarray}
in the first term where, in the rest frame of the incident electron, one has $\epsilon''p''_z=k'_z$, $\epsilon'p'_z=k'_z-k_z$, and
\begin{eqnarray}
&&\qquad\int{dp''_z\over2\pi}2\pi\delta(\epsilon''p''_z-\epsilon'p'_z+k'_z)2\pi\delta(\epsilon p_z-\epsilon''p''_z-k_z)=2\pi\delta(\epsilon p_z-\epsilon'p'_z+k'_z-k_z),\nonumber\\
&&\qquad\int{dp''_y\over2\pi}2\pi\delta(\epsilon''p''_y-\epsilon'p'_y+k'_y)2\pi\delta(\epsilon p_y-\epsilon''p''_y-k_y)e^{-ik'_x(\epsilon'p'_y+\epsilon''p''_y)/2eB}\,e^{ik_x(\epsilon''p''_y+\epsilon p_y)/2eB}\nonumber\\
&&\qquad\qquad=2\pi\delta(\epsilon p_y-k_y-\epsilon'p'_y+k'_y)e^{-ik'_x(\epsilon'p'_y+\epsilon p_y-k_y)/2eB}\,e^{ik_x(\epsilon'p'_y-k'_y+\epsilon p_y)/2eB},\nonumber\\
&&\qquad\qquad=2\pi\delta(\epsilon p_y-k_y-\epsilon'p'_y+k'_y)e^{i(k_x-k'_x)(\epsilon'p'_y+\epsilon p_y)/2eB}\,e^{i(k'_xk_y-k_xk'_y)/2eB},
\end{eqnarray}
in the second term where, in the rest frame of the incident electron, one has $\epsilon''p''_z=-k_z$, $\epsilon'p'_z=k'_z-k_z$.

One then has
\begin{eqnarray}
&&S^{(2)}_{fi}=(ie)^2\sum_q\sum_{q'}\sum_{\epsilon''}{\mu_0\over2V}{1\over\sqrt{\omega'\omega}}2\pi\delta(\epsilon'\varepsilon'+\omega-\epsilon\varepsilon-\omega'){1\over V^2eB}2\pi\delta(\epsilon p_z-\epsilon'p'_z+k'_z-k_z)\nonumber\\
&&\qquad\times2\pi\delta(\epsilon p_y+k'_y-\epsilon'p'_y-k_y)e^{i(k_x-k'_x)(\epsilon p_y+\epsilon'p'_y)/2eB}e^{*\mu}_M({\bf k})e^\nu_{M'}({\bf k}')\sum_{\sigma''=\pm}\sum_{n''=0}^\infty\,V(eB)^{1/2}
\nonumber\\
&&\qquad\times\left\{{e^{i({\bf k\,x\,k'})_z/2eB}[\Gamma^{\epsilon'\epsilon''}_{q'q''}({\bf k})]^\mu[\Gamma^{\epsilon\epsilon''}_{qq''}({\bf k}')]^{*\nu}\over\epsilon\varepsilon-\epsilon''\varepsilon''+\omega'}+{e^{-i({\bf k\,x\,k'})_z/2eB}[\Gamma^{\epsilon''\epsilon}_{q''q}({\bf k})]^\mu[\Gamma^{\epsilon''\epsilon'}_{q''q'}({\bf k}')]^{*\nu}\over\epsilon\varepsilon-\epsilon''\varepsilon''-\omega}\right\}.
\end{eqnarray}
The sums over the states of the incident and final particles (\cite{MP83}) are
\begin{equation}
\sum_q=\sum_{\sigma=\pm}\sum_{n=0}{\sqrt{eB}V\over L_yL_z},
\end{equation}
\begin{equation}
\sum_{q'}=\sum_{\sigma'=\pm}\sum_{n'=0}{\sqrt{eB}V\over(2\pi)^2}\int dp'_z\int dp'_y,
\end{equation}
respectively.  The integral over $p'_y$ gives
\begin{equation}
\qquad
\int{dp'_y\over2\pi}2\pi\delta(\epsilon p_y+k'_y-\epsilon'p'_y-k_y)e^{i(k_x-k'_x)(\epsilon p_y+\epsilon'p'_y)/2eB}=e^{i(k_x-k'_x)(k'_y-k_y)/2eB}
\end{equation}
for $\epsilon p_y=0$ (rest frame of incident electron) and $\epsilon=\epsilon'=+$.  The integral over $p'_z$ yields the implicit conservation of parallel momentum
$$\epsilon p_z-\epsilon'p'_z+k'_z-k_z=0,$$
which becomes
\begin{equation}
p'_z=k'_z-k_z=\omega'\cos\theta'-\omega\cos\theta,
\end{equation}
for $\epsilon p_z=0$ and $\epsilon=\epsilon'=+$.  With $L_x\equiv1/\sqrt{eB}$, one then has
\begin{eqnarray}
&&S^{(2)}_{fi}=(ie)^2\sum_{\sigma=\pm}\sum_{n=0}^\infty\sum_{\sigma'=\pm}\sum_{n'=0}^\infty\sum_{\sigma''=\pm}\sum_{n''=0}^\infty\sum_{\epsilon''=\pm}{\mu_0\over2V}{1\over\sqrt{\omega'\omega}}2\pi\delta(\epsilon'\varepsilon'+\omega-\epsilon\varepsilon-\omega')e^{i(k_x-k'_x)(k'_y-k_y)/2eB}\nonumber\\
&&\times e^{*\mu}_M({\bf k})e^\nu_{M'}({\bf k}')\left\{{e^{i({\bf k\,x\,k'})_z/2eB}[\Gamma^{\epsilon'\epsilon''}_{q'q''}({\bf k})]^\mu[\Gamma^{\epsilon\epsilon''}_{qq''}({\bf k}')]^{*\nu}\over\epsilon\varepsilon-\epsilon''\varepsilon''+\omega'}+{e^{-i({\bf k\,x\,k'})_z/2eB}[\Gamma^{\epsilon''\epsilon}_{q''q}({\bf k})]^\mu[\Gamma^{\epsilon''\epsilon'}_{q''q'}({\bf k}')]^{*\nu}\over\epsilon\varepsilon-\epsilon''\varepsilon''-\omega}\right\}.
\end{eqnarray}

Expressing the scattering matrix $S_{fi}$ in terms of a scattering amplitude $T_{fi}$,
\begin{equation}
S_{fi}=\delta_{fi}+i2\pi\delta(\epsilon'\varepsilon'+\omega-\epsilon\varepsilon-\omega')T_{fi},
\end{equation}
the probability per unit time of a transition is
\begin{equation}
w^{\epsilon'\epsilon}_{q'q}=V^22\pi\delta(\epsilon'\varepsilon'+\omega-\epsilon\varepsilon-\omega')|T_{fi}|^2.
\end{equation}
This leads to the probability for Compton scattering Eq.~(\ref{prob}).

\section{Decay rates}
\label{B-appendix}
The PC process of interest concerns the creation of an electron, $q'=n',p'_z,\sigma'$, and a positron, $q''=n'',p''_z,\sigma''$, by a photon with energy ${k'}^\mu=(\omega',{\bf k}')$, for which $p''^\mu=(\varepsilon'',{\bf p''})$, $p'^\mu=(\varepsilon',{\bf p'})$, with $\varepsilon''=\sqrt{m^2+{p''_z}^2+2n''eB}$, $\varepsilon'=\sqrt{m^2+(p'_z)^2+2n'eB}$ and where ${\bf e'_\perp}=(0,1,0)$ and ${\bf e'_\parallel}=(\cos\theta',0,-\sin\theta')$.  The $S$-matrix element is then
\begin{eqnarray}
&&S^{(1)}_{fi}=ie\sum_{q'}\sum_{q''}\int\,d^4x\,{\bar \psi}^+_{q'}(x)\Azp^\mu(x)\psi^-_{q''}(x),\nonumber\\
&&\qquad=ie{\sqrt{{\mu_0R_{M'}({\bf k}')\over\omega'_{M'}V}}}\sum_{q'}\sum_{q''}\int\,d^4x\,e^{\mu}_{M'}({\bf k}')
\{{\bar \psi}^+_{q'}({\bf x})e^{i\varepsilon't}\gamma_\mu e^{-i\omega't-i{\bf k'.x}}\psi^-_{q''}({\bf x})e^{i\varepsilon''t}\},
\end{eqnarray}
with $A'^\mu=\sqrt{\mu_0R_{M'}({\bf k}')/(\omega'_{M'}V)}e^{\mu}_{M'}(k')e^{-ik'_{M'}x}$.  The integral over time is trivial and yields $2\pi\delta(\varepsilon'+\varepsilon''-\omega')$, the conservation of energy equation.  With the definition
\begin{equation}
\qquad[\gamma^{+\,-}_{q'q''}({\bf -k'})]^\mu=\int d{\bf x}\,e^{i{\bf k'.x}}\,{\bar \psi}^+_{q'}({\bf x})\gamma_\mu\psi^-_{q''}({\bf x}),
\end{equation}
the spatial integral gives (\cite{MP83})
\begin{equation}
\qquad[\gamma^{+\,-}_{q'q''}({\bf k}')]^\mu={(2\pi)^2\over V\sqrt{eB}}\,e^{-ik'_x(p'_y-p''_y)/2eB}\delta(p''_y+p'_y-k'_y)\delta(p''_z+p'_z-k'_z)
[\Gamma^{+\,-}_{q'q''}({\bf -k'})]^\mu,
\end{equation}
with $[\Gamma^{+\,-}_{q'q''}({\bf -k'})]^\mu=[\Gamma^{-\,+}_{q''q'}({\bf k}')]^{*\mu}$ defined in Eq.~(\ref{Gammak}), so that one has
\begin{eqnarray}
&&S^{(1)}_{fi}=ie{\sqrt{\mu_0\over2\omega'V}}\sum_{q'}\sum_{q''}
{(2\pi)^2\over V\sqrt{eB}}\,e^{-ik'_x(p'_y-p''_y)/2eB}\delta(p''_y+p'_y-k'_y)\delta(p''_z+p'_z-k'_z)\nonumber\\
&&\qquad\qquad\qquad\times2\pi\delta(\varepsilon'+\varepsilon''-\omega')e^{\mu}_{M'}({\bf k}')[\Gamma^{-\,+}_{q''q'}({\bf k}')]^{*\mu}.
\end{eqnarray}
Now one has
\begin{eqnarray}
&&\sum_{q''}=\sum_{\sigma'',n''}\,{V\sqrt{eB}\over L_yL_z},\nonumber\\
&&\sum_{q'}=\sum_{\sigma',n'}\,{V\sqrt{eB}\over(2\pi)^2}\,\int\,dp'_z\,\int\,dp'_y,
\end{eqnarray}
so that the integral over $p'_y$ is
\begin{equation}
\int\,{dp'_y\over2\pi}\,2\pi\delta(p''_y+p'_y-k'_y)e^{-ik'_x(p'_y-p''_y)/2eB}=e^{-ik'_x(k_y-2p''_y)/2eB},
\label{dpy1}
\end{equation}
and the integral over $p'_z$ yields the implicit conservation of parallel momentum, $p''_z+p'_z=k'_z$.  Once again, the phase factor in Eq.~(\ref{dpy1}) disappears when $|S^{(1)}_{fi}|^2$ is taken.  The probability per unit time for pair creation is given by
\begin{eqnarray}
&&w^{M'}_{q''q'}=|S^{(1)}_{fi}|^2/T=V(2\pi)^4\delta^4(p_f-p_i)|T_{fi}|^2,\nonumber\\
&&\qquad\ ={\mu_0e^2\over2\omega'}\sum_{\sigma',n'}\sum_{\sigma'',n''}|{\bf e}_{M'\,\mu}({\bf k}')[\Gamma^{-\,+}_{q''q'}({\bf k}')]^{*\mu}|^2
\,2\pi\delta(\varepsilon'+\varepsilon''-\omega').
\end{eqnarray}
The pair creation rate is given by
\begin{equation}
R^{M'}_{q''q'}({\bf k}')={eB\over2\pi}\int{dp'_z\over2\pi}\,w^{M'}_{q''q'}({\bf k}'),
\end{equation}
where
\begin{equation}
\int\,dp'_z\,\delta(\varepsilon'+\varepsilon''-\omega')\,=\,{1\over|f'(p'_z)|}\,=\,{\varepsilon'\varepsilon''\over|\omega'p'_z-k'_z\varepsilon'|},
\end{equation}
with $f(p'_z)=\varepsilon'+\varepsilon''-\omega'$ implicit, and
\begin{eqnarray}
&&\varepsilon''=\sqrt{m^2+{k_z}^2+2n''eB},\nonumber\\
&&\varepsilon'=\sqrt{m^2+(k'_z-k_z)^2+2n'eB}.
\end{eqnarray}
This yields
\begin{equation}
R_{M'}=R^{M'}_{q''q'}={e^3B\mu_0\over4\pi\omega'}{\varepsilon'\varepsilon''\over|\omega'p'_z-k'_z\varepsilon'|}e^*_{M'\mu}({\bf k}')e_{M'\nu}({\bf k}')
\sum_{\sigma',n'}\sum_{\sigma'',n''}[\Gamma^{-\,+}_{q''q'}({\bf k}')]^\mu[\Gamma^{-\,+}_{q''q'}({\bf k}')]^{*\nu}.
\end{equation}

The pair creation rate is calculable for an electron and positron, with Landau quantum numbers $n'$ and $n''$ respectively, as long as $\omega'\ge2(\varepsilon''-k_z\cos\theta')/\sin^2\theta'$.  The pair creation rates for perpendicular and parallel polarized photons are explicitly
\begin{eqnarray}
&&R_\perp={e^3B\over4\omega'\varepsilon'_0}\,{1\over|\varepsilon''p'_z-k_z\varepsilon'|}\nonumber\\
&&\times\Big\{ \delta_{\sigma'+}\Big[(\varepsilon'_0-m)(\varepsilon'\varepsilon''-p'_zp''_z-m\varepsilon'_0)
(J^{n'}_{n''-n'-1}(x'))^2\nonumber\\
&&\ +(\varepsilon'_0+m)(\varepsilon'\varepsilon''-p'_zp''_z+m\varepsilon'_0)
(J^{n'-1}_{n''-n'+1}(x'))^2+2p_{n'}p_{n''}J^{n'}_{n''-n'-1}(x')J^{n'-1}_{n''-n'+1}(x')\Big]\nonumber\\
&&+\delta_{\sigma'-}\Big[(\varepsilon'_0+m)(\varepsilon'\varepsilon''-p'_zp''_z+m\varepsilon'_0)
(J^{n'}_{n''-n'-1}(x'))^2\nonumber\\
&&\ +(\varepsilon'_0-m)(\varepsilon'\varepsilon''-p'_zp''_z-m\varepsilon'_0)
(J^{n'-1}_{n''-n'+1}(x'))^2+2p_{n'}p_{n''}J^{n'}_{n''-n'-1}(x')J^{n'-1}_{n''-n'+1}(x')\Big]\Big\},
\end{eqnarray}
and
\begin{eqnarray}
&&R_\parallel={e^3B\over4\omega'\varepsilon'_0}\,{1\over|\varepsilon''p'_z-k_z\varepsilon'|}\nonumber\\
&&\times\Big\{ \delta_{\sigma'+}\Big[\cos^2\theta'\big\{ (\varepsilon'_0-m)(\varepsilon'\varepsilon''-p'_zp''_z-m\varepsilon'_0)
(J^{n'}_{n''-n'-1}(x'))^2\nonumber\\
&&+(\varepsilon'_0+m)(\varepsilon'\varepsilon''-p'_zp''_z+m\varepsilon'_0)
(J^{n'-1}_{n''-n'+1}(x'))^2-2p_{n'}p_{n''}J^{n'}_{n''-n'-1}(x')J^{n'-1}_{n''-n'+1}(x')\big\}\nonumber\\
&&+2\sin\theta'\cos\theta'\big\{ p_{n'}\big[(\varepsilon'_0p''_z-mp'_z)
J^{n'}_{n''-n'-1}(x')J^{n'-1}_{n''-n'}(x')+(\varepsilon'_0p''_z+mp'_z)
J^{n'-1}_{n''-n'+1}(x')J^{n'}_{n''-n'}(x')\big]\nonumber\\
&&-p_{n''}p'_z\big[(\varepsilon'_0-m)J^{n'}_{n''-n'-1}(x')J^{n'}_{n''-n'}(x')+
(\varepsilon'_0+m)J^{n'-1}_{n''-n'+1}(x')J^{n'-1}_{n''-n'}(x')\big]\big\}\nonumber\\
&&+\sin^2\theta'\big\{ (\varepsilon'_0+m)(\varepsilon'\varepsilon''+p'_zp''_z+m\varepsilon'_0)
(J^{n'-1}_{n''-n'}(x'))^2\nonumber\\
&&+(\varepsilon'_0-m)(\varepsilon'\varepsilon''+p'_zp''_z-m\varepsilon'_0)
(J^{n'}_{n''-n'}(x'))^2+2p_{n'}p_{n''}J^{n'-1}_{n''-n'}(x')J^{n'}_{n''-n'}(x')\big\}\Big]\nonumber\\
&&+\delta_{\sigma'-}\Big[\cos^2\theta'\big\{ (\varepsilon'_0+m)(\varepsilon'\varepsilon''-p'_zp''_z+m\varepsilon'_0)
(J^{n'}_{n''-n'-1}(x'))^2\nonumber\\
&&+(\varepsilon'_0-m)(\varepsilon'\varepsilon''-p'_zp''_z-m\varepsilon'_0)
(J^{n'-1}_{n''-n'+1}(x'))^2-2p_{n'}p_{n''}J^{n'}_{n''-n'-1}(x')J^{n'-1}_{n''-n'+1}(x')\big\}\nonumber\\
&&+2\sin\theta'\cos\theta'\big\{ p_{n'}\big[(\varepsilon'_0p''_z+mp'_z)
J^{n'}_{n''-n'-1}(x')J^{n'-1}_{n''-n'}(x')+(\varepsilon'_0p''_z-mp'_z)
J^{n'-1}_{n''-n'+1}(x')J^{n'}_{n''-n'}(x')\big]\nonumber\\
&&-p_{n''}p'_z\big[(\varepsilon'_0+m)J^{n'}_{n''-n'-1}(x')J^{n'}_{n''-n'}(x')+
(\varepsilon'_0-m)J^{n'-1}_{n''-n'+1}(x')J^{n'-1}_{n''-n'}(x')\big]\big\}\nonumber\\
&&+\sin^2\theta'\big\{ (\varepsilon'_0-m)(\varepsilon'\varepsilon''+p'_zp''_z-m\varepsilon'_0)
(J^{n'-1}_{n''-n'}(x'))^2\nonumber\\
&&+(\varepsilon'_0+m)(\varepsilon'\varepsilon''+p'_zp''_z+m\varepsilon'_0)
(J^{n'}_{n''-n'}(x'))^2+2p_{n'}p_{n''}J^{n'-1}_{n''-n'}(x')J^{n'}_{n''-n'}(x')\big\}\Big]\Big\}.
\end{eqnarray}
Only the rates with $n'=n''=0$ are of interest here, and these are
\begin{eqnarray}
&&R_\perp\ = 0,
\nonumber\\
&&R_\parallel\ ={e^3B\sin^2\theta'(\varepsilon'\varepsilon''+p'_zp''_z+m^2)e^{-x'}\over2\omega'|\omega'p'_z-k'_z\varepsilon'|}.
\label{Rat00}
\end{eqnarray}
If one sets $\theta'=\pi/2$, then one has $p'_z=-p''_z=\pm\half\sqrt{\omega'-4m^2}$ and $R_\parallel$ becomes
\begin{equation}
R_\parallel={e^3B\,m^2\,e^{-{\omega'}^2/2eB}\over{\omega'}^2|p'_z|},
\end{equation}
in agreement with Eq.~(6a) of \citet{DH83}, apart from an extra factor of 2 in their work which is accounted for once the integral of $\cos\theta$ is taken between $-1$ and $1$.

\end{document}